# Supervised Machine Learning for Eliciting Individual Demand

John A. Clithero[1], Jae Joon Lee[2], and Joshua Tasoff*[3]

[1]University of Oregon
[2]Stanford University
[3]Claremont Graduate University

February 1, 2021

**Abstract**

Direct elicitation, guided by theory, is the standard method for eliciting latent preferences. The canonical direct-elicitation approach for measuring individuals' valuations for goods is the Becker-DeGroot-Marschak procedure, which generates willingness-to-pay (WTP) values that are imprecise and systematically biased by understating valuations. We show that enhancing elicited WTP values with supervised machine learning (SML) can substantially improve estimates of peoples' out-of-sample purchase behavior. Furthermore, swapping WTP data with choice data generated from a simple task, two-alternative forced choice, leads to comparable performance. Combining all the data with the best-performing SML methods yields large improvements in predicting out-of-sample purchases. We quantify the benefit of using various SML methods in conjunction with using different types of data. Our results suggest that prices set by SML would increase revenue by 28% over using the stated WTP, with the same data.

---

*Clithero: Department of Marketing, Lundquist College of Business, 1208 University of Oregon, Eugene, OR, 97403 (email: clithero@uoregon.edu). Lee: The Digital Economy Lab at the Stanford Institute for Human-Centered Artificial Intelligence, Cordura Hall, 210 Panama Street, Stanford University, Stanford, CA 94305 (email: jaejoonl@stanford.edu) Tasoff: Department of Economic Sciences, Claremont Graduate University, 150 E 10th St, Claremont, CA, 91711 (email:joshua.tasoff@cgu.edu). We thank Jin Xu for valuable research assistance. We would like to thank participants at seminars at CMU, University of San Francisco, Case Western Reserve, Colegio de Mexico, Western Virginia University, The Wharton School, Karlsruhe Institute of Technology, Claremont Graduate University, USC, the WEAI Conference, and the ASSA Annual Meeting.

# 1 Introduction

Many economists seek to uncover aspects of consumers' latent preferences, such as their demand for a good, their tolerance of risk, or the way they discount the future. Direct elicitations incentivize individuals to make choices in such a way that they directly reveal their latent preferences. These direct elicitations could be either structure-free, such as in the Becker-DeGroot-Marschack (BDM) procedure, in which a person is incentivized to truthfully state her reservation price for a good (Becker, DeGroot and Marschak, 1964), or structural, such as fitting choices between lotteries to a parametric model of a consumer's utility function to yield a risk aversion coefficient (for example, Holt and Laury, 2002; Eckel and Grossman, 2008; Andersen et al., 2008; Anderson and Mellor, 2009). In both cases, theory provides a mapping between the data and preferences.

Direct elicitation is granted grace from a strong conceptual framework, but it often works better in theory than in practice. We present an alternative approach that can enhance direct elicitation using supervised machine learning (SML). We borrow the term from the computer science and statistics literature (James et al., 2015), but it is a concept familiar to all economists (Varian, 2014). SML is simply the set of statistical methods that relate input variables to output variables. Regression in its many forms is a kind of SML.[1] In this paper we take an SML approach to recovering consumers' latent valuations for goods. The mapping between data and latent preferences is no longer set by theory as with a direct elicitation, but is instead uncovered statistically. Attempts to address the problems with the BDM procedure have thus far focused on fixing the front end by modifying the elicitation procedure to be simpler or more intuitive (for example, Wang et al., 2007; Miller et al., 2011; de Meza and Reyniers, 2013; Mazar et al., 2014). The SML approach is to fix the problem on the back end by estimating the statistical relationship between the responses and the outcome behavior.

As a proof of concept, we apply SML to estimate a consumer's demand for a good. We show that SML outperforms raw BDM predictions of a person's purchase behavior. This is an important test, as the BDM is arguably the gold standard for eliciting WTP, the measured reservation value, for goods. Our paper presents a series of consumer choice tasks designed explicitly for an out-of-sample comparison of these approaches. Subjects in our experiment engage in three tasks involving choices over food items. The first is a task designed to elicit WTP (BDM-Task). The second is a binary choice task, which we refer to as a two-alternative forced choice task (2AFC-Task). The third task is a simple decision on whether to buy a good at a fixed price (Buy-Task). We thus have an outcome measure that best represents the kind of decisions that consumers actually make in the marketplace.

[1] In contrast, factor analysis, clustering, and other statistical methods that do not have an outcome variable are forms of *unsupervised* learning.

We find that BDM measures enhanced by SML predict purchase behavior better than the BDM measures alone, even with small amounts of data and using simple statistical methods such as logit. This is partially explained by the fact that the BDM has a downward bias in our sample. For example, subjects who state a WTP of $\$x$ will often still purchase the good when the price is $\$x + \$0.25$. However, this is not the whole story as there are considerable gains in performance as more sophisticated methods are used, such as random forest, suggesting that there are correlations in the choice data that can be leveraged to improve prediction. We also show that more data is not a substitute for better statistical methods. As sample size increases, low-dimension methods (e.g., logit) plateau at a higher mean squared error (MSE) relative to high-dimension methods (e.g., lasso and random forest). With random forest, we find that WTP data and 2AFC data perform about equally in predicting buy-decisions, out-of-sample, using a combined between- and within-subject analysis.

This is a notable result given that the 2AFC choice data contain only *ordinal* information about subjects' preferences between the various snacks and does not contain any *cardinal* information about the money-metric intensity of those preferences. WTP data contains exactly that *cardinal* information. Initially, the effectiveness of SML with 2AFC data may appear counterintuitive. However, there is a parsimonious explanation. An array of 2AFC data can be used to construct an individual's ranking of the goods. When paired with the outcome variable of purchases decisions, these rankings can be linked to prices. For example, if a person prefers a granola bar to fruit juice and is willing to buy the fruit juice at \$1.25, it is also likely that the person will purchase the granola bar at \$1.25. We use SML to apply this kind of data and logic at scale.

Interestingly, SML with 2AFC suffers considerable decreases in accuracy when we restrict prediction to when we train the model on one sample of subjects and predict purchases on an entirely different sample of subjects. In contrast, SML with WTP suffers only minor decreases in accuracy when prediction is evaluated in this way. This suggests that patterns of preference within consumers are not effective at predicting the intensity of preference across consumers. For example, two people may have the exact same rankings of the food items but the first person might be very hungry and have high demand for all items, while the second person may be satiated and have low demand for all items. The WTP data can reflect this but the 2AFC data cannot.

Our best predictions, though, come from using both data together, suggesting the existence of non-overlapping information. Combining data from different elicitation methods improves performance. This result holds for both lasso and random forest.

We also show that revenue-maximizing prices predicted from our best SML models yield important deviations from a consumer's WTP. The two values are correlated, but the average absolute deviation is \$0.60. Given that the average WTP of the good in our experiment is \$2.03, this implies a 29.4% deviation in optimal pricing. Using our best SML model, we predict that our SML pricing would yield revenues 27.6% higher than directly setting prices to the WTP value, without using any additional data.

We view the main contributions of this paper as two-fold. First, our results draw attention to the surprisingly neglected approach of using SML to estimate latent preferences. Second, on a practical level, we show that SML can improve upon the existing canonical direct elicitation method, the BDM. Thus, we view our exercise as a necessary proof-of-concept that the SML approach to estimating latent preferences can be effective. While our particular elicitations and statistical methods are not necessarily the global optimum for estimating individuals' valuations for goods, our framework offers a template for future work in both applied and theoretical areas.

Our paper is structured as follows. Section 2 discusses additional motivation and conceptual background. Section 3 presents our experimental design and Section 4 our empirical strategy. Sections 5, 6, and 7 report our results on direct elicitation, SML, and optimizing data use. In Section 8 we quantify the benefits of using SML. Section 9 includes a brief discussion and Section 10 concludes.

## 2 Eliciting Willingness to Pay and Predicting Behavior

Accurately forecasting consumer demand is a fundamental aim for all firms. There exist many innovations in both "indirect" and "direct" methods for elicitation (Miller et al., 2011), in both real (Ding, 2007) and hypothetical choice scenarios (Schmidt and Bijmolt, 2020). Our contribution builds on arguably the most popular direct elicitation method, the BDM. This section details some limitations of the BDM, why other simple choice data can help fill the gap, and why SML can make a meaningful contribution to building better predictions of individual consumer demand.

The BDM method has been hugely influential across marketing and economics. As of January 2021, Google Scholar lists more than 2900 citations for the original paper (Becker et al., 1964). However, it has several well-established limitations. Consider the following three issues. First, the BDM produces measures that are systematically biased. Lehman (2015) and Müller and Voigt (2010) find that WTP from the BDM are overstated and Berry et al. (2020); Wertenbroch and Skiera (2002); Noussair et al. (2004); Kaas and Ruprecht (2006) find that WTP from the BDM are understated. We show additional evidence for this understatement. Second, direct elicitation requires the production of specific types of choice data that pin down the latent parameters of a formal theory. For the case of measuring people's values for goods, many researchers have observed that the complexity of the BDM confuses subjects, which results in noisy and biased responses (see e.g., Cason and Plott, 2014). Past research has found that procedural variations of the BDM that theoretically should have no effect on people's elicited responses instead have a profound effect (Urbancic, 2011; Mazar, Kőszegi and Ariely, 2014; Tymula, Woelbert and Glimcher, 2016). Third, practitioners of direct elicitation often (though not always, see Wang et al., 2007) just-identify their parameter of interest as that is sufficient according the the conceptual framework of direct elicitation. However, the presence of measurement error, be it through confusion, random shocks to utility, or literal trembling hands can lead to severe mis-inference (Gillen, Snowberg and Yariv, 2019). In our data, we observe both purchases at prices that are above a person's elicited WTP and non-purchases at prices that are below a person's elicited WTP. Such consumer behavior is difficult to reconcile with consumer theory that does not build some flexibility into individual demand around a directly elicited WTP.

We believe SML offers a unique solution to the three issues with the BDM that we mentioned above. First, if WTP values are systematically biased – as they are in our data – SML can de-bias these values by using a loss function that treats positive and negative errors symmetrically. In other words, by identifying any bias in WTP within the training sample, SML should not have such a bias in out-of-sample prediction. Second, SML can obviate the participant confusion sometimes caused by complex elicitation mechanisms, by instead relying on data generated from simple tasks. Input variables for SML are far less constrained than in direct elicitation. Potentially any kind of data (i.e., non-choice data) could be used by SML as long as the data is a good predictor for the outcome, including the data from a direct elicitation. This relaxation on data inputs allows for innovative data generation. In this paper, we estimate valuations using direct elicitation data produced using the BDM, and we illustrate how an SML approach can rely on simpler choice tasks and even non-choice data, such as response times (RT). Third, the flexibility conferred by the SML approach also allows the researcher to leverage repeat observations, reducing measurement error.[2]

Some hurdles that all applications of SML face are worth mentioning. First, outcome data is required. The BDM can elicit a person's WTP without the need for observing any real purchase decisions. In contrast, SML needs that outcome data for model training. Second, direct elicitation, by assumption, identifies the latent preference with a single observation. Because SML is a statistical method it needs multiple observations, usually many, to generate accurate estimates. There is an additional reason why SML has been largely avoided as an approach for measuring latent preferences in favor of direct elicitation methods. Historically, a challenge facing SML was "knowledge discovery"; how does one find the statistical relationship between input variables and outcomes without overfitting? Advances in machine learning have effectively solved this problem (James et al., 2015). Algorithms search and test many specifications using brute-force computation and then settle on the best parameter values, based on out-of-sample prediction. Empirically, many popular algorithms are now effective at predicting outcomes in real-world social science data (see Lazer et al., 2009; Varian, 2014; Hagen et al., 2020, for some examples).

Our paper adds to what is a growing literature applying machine learning to help predict consumer behavior, both in marketing and economics. In consumer behavior, SML has been used to understand heterogeneous responses to marketing images (Matz et al., 2019), customer retention (Ascarza, 2018), and consumer finances (Netzer et al., 2019), among others. In the domain of price-setting, Bodoh-Creed et al. (2019) use SML to better predict variation in prices for homogenous products. For our purposes, there are also several recent papers using SML in economics. The closest to ours is Peysakhovich and Naecker (2017), who compare structural models of risk aversion and ambiguity aversion to SML predictions in order to evaluate the validity of the structural models. They find that structural models of risk aversion perform as well as SML but structural models of ambiguity aversion perform far worse. Our paper makes a different point: we provide a proof-of-concept for recovering latent preferences using SML. Bernheim et al. (2013) use SML and non-choice data to predict the behavior of populations in aggregate. They generate voluminous data on preferences for goods using non-choice survey data and then use this to predict aggregate demand for the good at a given price. Naecker (2015) applies this method to predict organ-donation registration. Smith et al. (2014) use SML on brain imaging data to make out-of-sample predictions on human choices. Our paper is in a similar spirit to these papers, but with a different practical goal: we are predicting individual behavior instead of aggregate behavior, and we use various kinds of choice data instead of explicitly non-choice data. Finally, Camerer, Nave and Smith (2018) apply SML to predict outcomes in bargaining experiments.

[2] In principle, one can repeat a direct elicitation multiple times and then average the measures together. This is uncommon and could be considered the first step to an unsupervised learning approach (i.e. statistically processing the measures without the use of outcome data).

A common theme across many of these SML applications is that there are aspects of a consumer's preferences unobservable to the researcher. Direct elicitation may do a good job of identifying consumer WTP, but constructing how this maps to purchase decisions requires a mapping that is unseen. We believe SML provides one path for constructing such individual demand curves. The next section details our framework for testing this claim.

# 3 Experiment Design

Subjects engaged in three choice tasks for 20 different food items.[3] Food items have the advantage of being commonplace and therefore easy to assess. There were 20 BDM trials, 190 2AFC trials, and 80 Buy trials, as described below. Though this may sound like a large number of tasks, each trial takes only a few seconds. All the tasks combined took less than 30 minutes on average (excluding waiting time, eating time, etc). Subjects were told that one trial would be randomly selected to count and they would get the food item depending on their choice. Prior to engaging in the tasks, two of the 20 food items were randomly selected to be bonus "silver" and "gold" items. Subjects saw 20 covered cards on the screen and selected two. The cards were flipped over to reveal the bonus items. Subjects received bonus payments of $2 for obtaining their silver item and $4 for obtaining their gold item. The purpose of these bonus items was twofold. First, it provided a check on subjects' preferences, as the monetary bonuses should increase WTP (it did). Second, it created greater variation in peoples' valuations for the food items, making the prediction task harder.

The first task elicited the WTP for each of the 20 items (BDM-Task). The task was similar to a multiple price list (MPL) version of a BDM mechanism. On the computer screen, each subject is asked to choose her WTP for each of 20 items from $0 to $5.75 in increments of $0.25. To help reduce concerns about participant confusion (Cason and Plott, 2014), we provided detailed instructions about the BDM and also asked subjects to complete a quiz containing five questions before starting the task. Copies of experimental instructions and the quiz are provided in the Appendix.

The second task was two-alternative forced choice (2AFC-Task). On each trial, subjects were shown a picture of two items: one placed on the left side of the computer screen and one on the right. Subjects had to decide which of the two items they would prefer to eat at the end of the experiment by clicking left or right. Subjects were free to take as long as they needed to make their decision. Subjects were asked to make a choice between all possible pairs of items, so each subject had 190 binary choices. In addition to each choice itself, each subject's RT was collected for each trial. We chose the 2AFC-Task for several reasons. First, we thought it would be one of the simplest most easily understood ways to get real choice data. Second, response time in the 2AFC-Task has been shown to be predictive of future choice (Clithero, 2018). Third, the 2AFC-Task, as a simple binary choice task, represents arguably the most common form of a choice task in all of behavioral science.

[3]We conducted a pilot study to determine the final set of 20 items from a candidate set of 40 foods. We elicited the WTP of 40 items and selected the 20 items which showed the highest variation in WTP across individuals. The final list of 20 food items is available in Table A1 in the Appendix.

In the third task, subjects faced a single food item and a posted price and had to decide whether to buy each snack or not at the posted price (Buy-Task). Each snack was shown four times with four different prices. One of the four prices was the stated WTP in the first main task, and the other three prices were randomly selected from uniform distributions over the low, medium, and high supports $[0.25, 1.00]$, $[1.25, 2.00]$ and $[2.25, 5.75]$ respectively, each segmented in 25-cent increments. The order of the items and prices was presented randomly.[4] The Buy-Task is arguably most similar to what consumers actually face: a single good listed at a single price with the option to buy or not. The behavior in the Buy-Task is what we are trying to predict. Figure 1 shows screenshots of the tasks. We use various subsets of data from the BDM-Task and the 2AFC-Task, fed into a variety of models to predict behavior in the Buy-Task. Details about the statistical methods are explained below in Section 4.

Participants were recruited through the maintained subject pool at a western university in the United States from March to August in 2017. Each subject was asked to abstain from eating for 3 hours prior to the experiment. Subjects had to be fluent in English and have no dietary allergies or restrictions that would prevent them from consuming common snack foods. Upon arrival at the lab, all participants were consented and provided paper print-outs of the instructions. Instructions were then read out loud by the experimenter.

Each subject received the participation fee of $20, but the actual amount of cash received varied depending on her choices during the experiment and the trial randomly selected to count. In total, 55 subjects participated in the study over eight sessions. Subjects who received a food item at the end were required to take at least a single bite to prevent resale in a secondary market.[5] The experiment lasted approximately one hour.

[4]The order of the three tasks was kept constant over the course of the experiment. The Buy-Task was placed after the WTP-Task so that we could use the elicited price as one of the prices.

[5]Anecdotally, the authors observed that virtually all participants consumed the entire food item prior to

Figure 1: Tasks Overview and Models

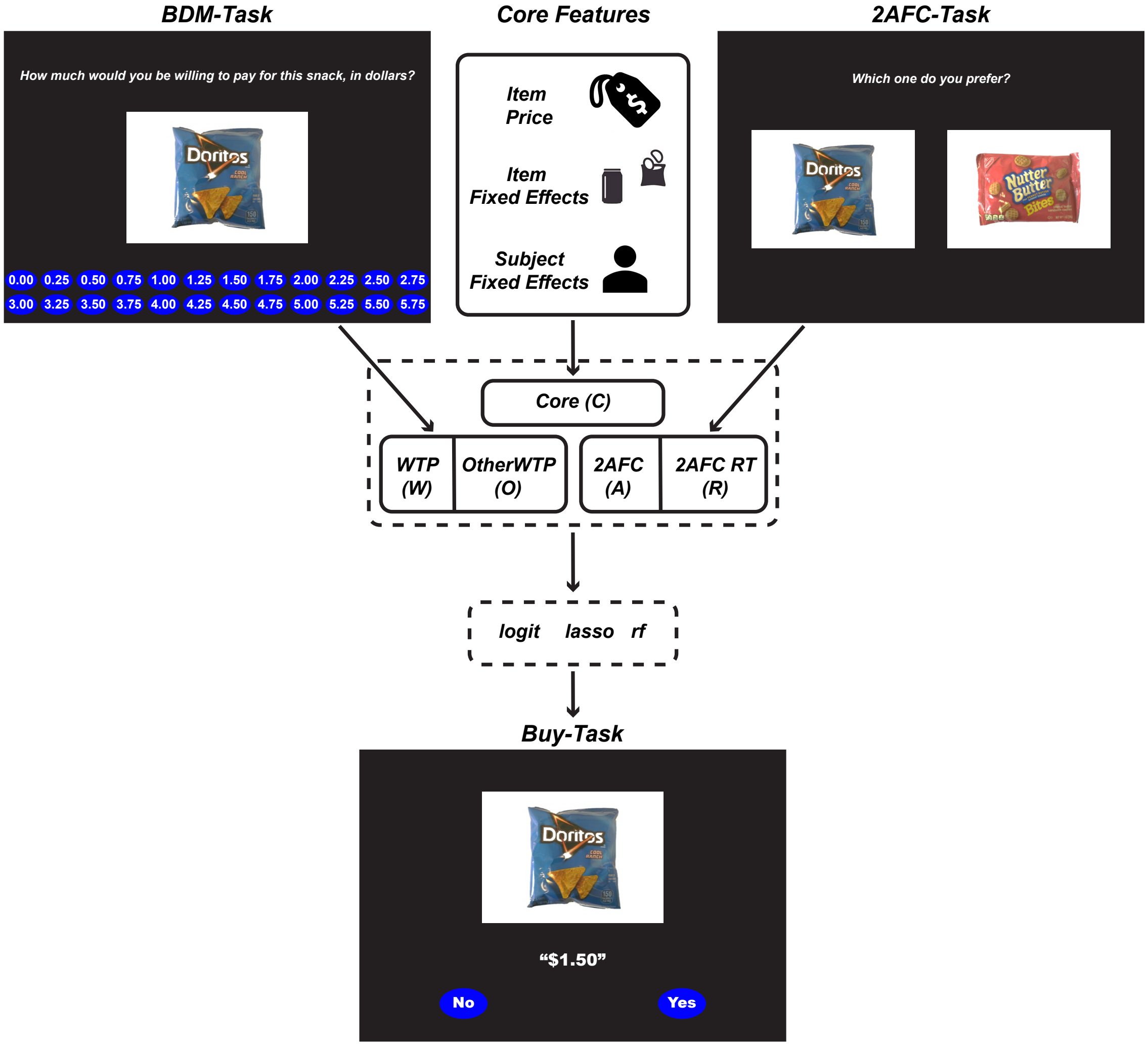

Notes: Summary of the empirical strategy. Three different tasks were completed by subjects: BDM-Task, 2AFC-Task, and Buy-Task. Two different types of data from the first two tasks, WTP and 2AFC, along with Core features (top row), are used to construct feature spaces (second row). These feature spaces are subsequently inputted into one of three algorithms (third row) to make out-of-sample predictions on purchase decisions (bottom row).

# 4 Empirical Strategy for Prediction

## 4.1 Basic Comparison

Our main purpose is to compare our direct elicitation method, the BDM, to SML methods. Thus the prediction of the BDM is our direct elicitation benchmark. We define the prediction of the BDM to be buy with probability 1 if the price is less than the WTP, and buy with probability 0 if the price is strictly greater than the WTP. If the price exactly equals the WTP the person is indifferent and buys at probability $q$. We set $q$ to the empirical average purchase frequency when $p =$WTP.

There are many SML methods available and two different data sets. In addition to assessing the performance of SML methods relative to the BDM, we also wish to characterize the comparative performance of the various SML methods, and the comparative performance of the two different data formats. We begin by smoothing the BDM prediction, allowing noise in the WTP via a logit according to equation (1):

$$\log\left(\frac{buy_{ijt}}{1-buy_{ijt}}\right) = \beta_0 + \beta_1 p_{ijt} + \beta_2 WTP_{ij} + \sum_{k=2}^{20} 1\{k=j\}\cdot(\gamma_k + \delta_k p_{ijt} + \zeta_k WTP_{ij}) + \sum_{k=2}^{55} 1\{k=i\}\cdot(\eta_k + \iota_k p_{ijt} + \kappa_k WTP_{ij}) + \epsilon_{ijt}, \quad (1)$$

which constructs a probability the individual will purchase the item at the stated price, $P(buy_{ijt})$. The buy-decision for person $i$ on item $j$ on trial $t$ (where $t = 1, 2, 3, 4$) is a function of a constant, the price $p_{ijt}$, person $i$'s WTP for item $j$, item fixed effects $\gamma_k$, and item fixed-effect interactions with the price and WTP ($\delta_k$ and $\zeta_k$, respectively), subject fixed effects $\eta_k$, subject fixed-effect interactions with the price and WTP ($\iota_k$ and $\kappa_k$), and an error term, $\epsilon_{ijt}$. An alternative baseline comparison built around consumer surplus that performs similarly is outlined in Appendix B.

This regression model does not make full use of all the WTP data available. Instead of merely including the WTP of the item in question we could also include the full vector of WTPs across all 20 items. Equation (2) is the same as Equation (1) with two additions. First, on the second line, we include the WTP for the other 19 food items, indexed with $k$. Second, on the third line, the double summation term is item fixed-effects interacted with the WTP for the other 19 food items. This allows for the WTP of a given good $k$ to have a leaving the experiment. This allows WTP measures to be interpreted in terms of a full item.

different effect on each good $j$:

$$
\begin{aligned}
buy_{ijt} =& \beta_0 + \beta_1 p_{ijt} + \beta_2 WTP_{ij} \\
&+ \sum_{k=2}^{20} \left(\theta_k WTP_{ik} + 1\{k=j\} \cdot (\gamma_k + \delta_k p_{ijt} + \zeta_k WTP_{ij})\right) \\
&+ \sum_{l=1}^{20} \sum_{k=2}^{20} \lambda_{k,l} 1\{k=j\} \cdot WTP_{il} \\
&+ \sum_{k=2}^{55} 1\{k=i\} \cdot (\eta_k + \iota_k p_{ijt} + \kappa_k WTP_{ij}) + \epsilon_{ijt}. \qquad (2)
\end{aligned}
$$

We will refer to the first model as the logit(WTP), since it only uses the WTP for the item in question, and logit(WTP+OtherWTP) for the more complete model. The latter model is not obviously better than the former for prediction because it could overfit the data.

## 4.2 High-Dimensional Methods

"High-dimensional" methods are methods that are well-suited to analyzing data with a large number of independent variables. We use two off-the-shelf methods that are widely adopted in the literature (all estimation was conducted using freely-available packages in R). We will describe the basic intuition and mechanics here. More detailed treatments can be found in textbooks such as Hastie et al. (2009) and James et al. (2015). The first method is a logistic regression with a least absolute shrinkage and selection operator (lasso) penalty (Tibshirani, 1996). The lasso algorithm first normalizes all the independent variables. Then a regression is run, but for each independent variable included in the regression there is a marginal penalty applied to the objective function (e.g. mean squared error or binomial deviance, depending upon outcome variable of interest), for each unit that a coefficient is away from zero. Because the marginal loss of increasing the magnitude of a coefficient is constant, coefficients on variables that do not reduce the error term enough lead to a corner solution, yielding coefficients of zero. This method is well-suited for high-dimensional data as it does not include variables in the model that do not substantially improve prediction. Since our outcome is dichotomous, the lasso we use is a penalized logit in which the objective is to minimize binomial deviance. Lasso is already popular in economics (Mullainathan and Spiess, 2017) and is an intuitive modification to a logit prediction.

Tree methods take a different approach. A tree is a partition of the data set that assigns the mean of the outcome within the partition to be the predicted value for all observations in the partition. In other words, it is a regression run on partition indicators and nothing else. The partitions are built using a myopic algorithm that first selects an independent variable and a value, then creates an indicator partitioning the data set into all observations below this value and all observations above this value. This is repeated for every possible division and every independent variable. The indicator that improves the objective function by the most is selected to be the first "node" in the tree. The process is then repeated several times to create the tree.

A random forest is an algorithm that builds many trees on the same data set and averages the predictions together to reduce variance. There are two randomizations in the generation of the trees. First, many bootstrap samples using the full sample size are produced. A tree will be produced for each bootstrap sample. Second, only a random subset of independent variables will be included in the generation of branches at each node of each tree. This restriction de-correlates the trees reducing the overall variance of the estimator (James et al., 2015). We use random forest because is a widely used method, and has a record for high performance (Athey and Imbens, 2019).

Each of these methods have tuning parameters. Lasso has a penalty parameter. Random forest has parameters for the number of nodes in the tree, the number of trees, and the size of the subset of variables to be considered at each node. We optimize these parameters according to the conventional methods in machine learning, which involves different tuning approaches for the different methods. For lasso we used an approach called 10-fold cross-validation. First we split the sample randomly into a 90% training set and a 10% test set. The training set is then randomly split into 10-folds (i.e., subsets). A parameter set is chosen, and the model is estimated on 9 of the 10 folds with the loss function (e.g. MSE or binomial deviance) measured out-of-sample on the remaining validation fold. This is then repeated 10 times, each time rotating the validation fold. The loss averaged over the 10 folds is a performance metric for that given parameter set. The process is repeated iteratively over the entire parameter space. The parameter set that gives the least loss is selected. The model with the optimized tuning parameters is then re-estimated using the full training set and evaluated using out-of-sample loss on the test set, as detailed in the next section. Random forest, the most computationally intensive algorithm we used, employs a different tuning procedure. Random forest offers a straightforward way to estimate out-of-sample performance without cross-validation. Each bootstrap sample omits approximately 1/3 of observations, referred to as out-of-bag (OOB) observations (James et al., 2015). One can predict a given observation by using the average of all trees for which the observation is OOB. The error is referred to as the OOB error. The algorithm then selects the parameter set to minimize the OOB loss (e.g. MSE or binomial deviance).

## 4.3 Performance Metrics

We evaluate performance based on out-of-sample prediction. As described above, for each analysis we have a randomly selected training set and test set. In our main analysis, we randomize at the observation level, where each observation is a buy-decision (four buy-decisions per subject-item). We reserve 440 observations (10%) of our data as the test set. The 440 observations were drawn randomly using stratified sampling. We divided the data into quintiles, based on subject's percentage of purchase decisions. An equal number of observations were randomly drawn from each of these five bins. Due to the potential variance generated from a single random test set, we repeat the process $N$ (here, $N = 50$) times. This repetition helps smooth out the prediction and ensures differences in performance are not attributable to particularities of one test set.

In the main body of the paper we use mean squared error (MSE) as our performance metric due to its widespread use. The MSE across the $N$ test sets is then as follows:

$$MSE = \frac{1}{N} \sum_{n} \sum_{i,j,t} \left( P(buy_{ijt}) - \hat{P}(buy_{ijt}) \right)^2, \tag{3}$$

where $P(buy_{ijt})$ is the observed decision (i.e., ether 1 for "Yes" or 0 for "No") and $\hat{P}(buy_{ijt})$ is the predicted choice probability from SML. Here, each MSE test set $n$ effectively captures the mean squared error across all subjects, items and prices. We then average across the $N$ randomly-constructed test sets to obtain our final measure of prediction performance.

Two other measures that appear in the literature are binomial deviance and "area under the curve" (AUC). Binomial deviance is the negative of the binomial log-likelihood. In a logit regression, the objective is to minimize this quantity. The AUC is a metric that summarizes the performance of a binary classifier and is often used in the machine learning literature. A perfect classifier will have an AUC equal to 1, whereas a random classifier would have an AUC equal to 0.5. In our analyses, we consider MSE and AUC; both metrics lead to similar results. MSE has arguably the most desirable properties as binomial deviance becomes infinite when a model makes a prediction with certainty and is wrong (as the BDM does), and AUC requires large amounts of data. For completeness, we include AUC results in Appendix B.

## 4.4 Data and Features

To perform our prediction exercise, we need to create the explanatory variables, a step commonly called feature construction. While modern SML methods have taken the art out of specification search and replaced it with an effective systematic approach, there remains a need for the researcher to generate the explanatory variables. The SML literature refers to these explanatory variables as features. We construct several sets of features, each stemming from one of our sources of data. We then leverage this modular design to understand how different kinds of data contribute to our ultimate prediction exercise. A summary of these feature modules is in Table 1 and a full list of all features can be found in Appendix A.2.

The first set of features for us is a set of "Core" features that will be used in all predictive models. The features are constructed using price variables, fixed effects for subjects, and fixed effects for items (i.e. indicators for subjects and items). These features can also serve as a baseline for prediction, as a practitioner could construct this set of features without any additional data. In other words, if we want to predict purchase behavior in our Buy-Task, these are features we can use without considering any of our WTP-Task or 2AFC-Task data.

As Section 4.1 explains, we make use of the WTP data set in two ways to construct features. "WTP" (W) models use data which only uses the WTP of the item in question in the analysis including: a polynomial expansion of WTP and the price, item fixed effects, interactions between WTP and the item indicators. The more expansive "WTP+OtherWTP" (WO) models include all those features as well as interactions between every WTP and every item indicator. These models include the full 20 item vector of WTP values in each regression. We begin by running simple logit models to predict the buy decisions as expressed in Equation (1) and Equation (2) giving us the models logit(W) and logit(WO). After logit, we run random forest on the W and WO feature spaces. Comparing logit(W) to the random forest model, rf(W), tests whether there is a gain in performance from using high-dimensional methods but with low-dimensional data. Comparing logit(WO) to random forest with the full feature set, rf(WO), tests whether there is a gain in performance using the high-dimensional method with high-dimensional data.[6]

The second data set is from the 2AFC-Task. We have both the choice data and RT. On this data set we run the same algorithms, logit, lasso, and random forest each with and without RT data. The 2AFC data require more careful feature construction to have data that is in a form that the algorithms can use effectively. The "2AFC" (A) models add 211 different features in our regressions based on the binary choice data. Our approach is to create item fixed effects and dummy variables for an item being chosen over a specific different item. For example, a dummy if the item in question was chosen over Godiva Dark Chocolate, another dummy if the item in question was chosen over KIND Nuts & Spices, and so forth. We also construct variables for the frequency with which the item in question was chosen.

[6]Our point here is to measure how much improvement there is from using high-dimensional data *only*, high-dimensional methods *only*, and both *simultaneously*. For exposition, we omit the lasso models using WTP as they perform worse than random forest in this context. With the 2AFC data, which is higher dimension than the WTP, we devote more time to comparing the high-dimension methods, lasso, and random forest, to each other.

As mentioned earlier, we recorded RT data on each 2AFC trial. The existing literature suggests shorter RT should be correlated with a stronger preference for the chosen good relative to the unchosen good (Clithero, 2018; Krajbich et al., 2010; Philiastides and Ratcliff, 2013) and that RT data can increase predictive performance in certain choice environments (Clithero, 2018). We add another 87 features when we include the RT data (R). The main type of feature in consideration is an interaction term between dummies for whether the item in question was chosen over a specific item (e.g., it takes the value of one when the item in question was chosen over KIND Nuts & Spices) with the RT. In principle, RT data could mitigate the primary weakness of 2AFC data, which is a lack of data on preference intensity.

All combined Core, WTP, and 2AFC data lead to 836 features without RT and 923 features with RT. This partitioning of our feature space is an effort to combat a common criticism of SML, that it often does not have interpretable results. By creating modules that come from different kinds of data, we can isolate where there is added predicted value of the data.

Table 1: Summary of Feature Models

| Groups of Features | Total Features | Abbreviation |
|---|---|---|
| Core | 149 | C |
| Core + WTP | 225 | W |
| Core + WTP + OtherWTP | 625 | WO |
| Core + 2AFC | 360 | A |
| Core + 2AFC + RT | 447 | AR |
| Core + WTP + 2AFC | 436 | WA |
| Core + WTP + OtherWTP + 2AFC | 836 | WOA |
| Core + WTP + OtherWTP + 2AFC + RT | 923 | WOAR |

## 4.5 Statistical Models and Research Questions

Our empirical strategy is summarized in Figure 1 and Table 1. Figure 1 diagrams the tasks, data, and how the data feeds into the various statistical models. As discussed, we have several baseline prediction methods to use as benchmarks. We have the BDM prediction, which will not require any model fitting. As another simple prediction metric, we have the empirical purchase frequency of our sample, which can be used as the probability an individual purchases the item on each trial (which we will refer to as "Prob(Buy)"). Finally, we have the set of Core features, which we run in each of the algorithms used. We use logit, lasso, and random forest.

The nature of our dataset also allows us to vary the sample size of the data used to train algorithms. We are thus able to investigate if and how this impacts prediction performance. We are also interested in the extent to which high-dimensional methods (here, lasso and random forest) can improve predictions. We can compare algorithms with the same data as inputs. For example, comparing rf(WO) to logit(WO) reveals the improvement from using a high-dimensional method over logit.

As we have two different kinds of data, direct elicitation (WTP) and binary choice (2AFC), we can compare the performance of these data using the various feature space modules summarized in Table 1. We can also look within the 2AFC data and determine if, given all of the other data we have, RT data can be of additional predictive value for our prediction exercise.

Finally, we also wish to know whether the 2AFC data and the WTP data are redundant or if there is unique information in each data set that would imply improved performance by using both? For this question we run models using feature space combinations of both, WA and WOA. Comparing rf(WOA) to rf(WO) and rf(A) reveals the improvement from combining WTP and 2AFC data together.

To summarize, the analysis will allow us to answer several important questions:

1. What is the degree of bias in WTP using the BDM to measure reservation value? This can be done by measuring the predicted probability of purchase when the price is exactly at the WTP.

2. How does algorithm performance vary with sample size? Algorithms with more features may require higher sample sizes for good prediction. We can observe for what sample sizes a high-dimensional algorithm outperforms a low-dimensional model.

3. Which algorithm performs best with our data? We compare three different algorithms, as outlined in Section 4.2.
4. Which data best predicts purchases? We compare all feature spaces within a given algorithm.
5. Are WTP and 2AFC data redundant? We can test this by constructing feature sets with various combinations of WTP and/or 2AFC data.

The next three sections answer these questions directly.

# 5 Direct Elicitation

We begin our analysis with a look at the data. High quality prediction is only interesting if there is a high degree of variation in the behavior being predicted. For the prediction task to be interesting, we want the data set to have a high degree of variation in WTP within-item and within-consumer. If everyone values a Hershey's Bar at exactly $1.25, prediction is trivial. Likewise, if demand for all goods can be easily characterized by a one-dimensional parameter, such as hunger, our elicitation devolves to just measuring hunger. A more challenging task is one in which there is heterogeneity in both hunger and tastes.

Figure 2(a) shows the distribution of WTP by item. There is a high degree of heterogeneity across items, with average WTP ranging from $1.10 for a 12 oz Coke (CK) to $2.99 for Naked Mango Juice (NM). There is also considerable heterogeneity within item. The mean WTP is $2.03 and the standard deviation is $1.27. Figure 2(b) shows the heterogeneity across individuals. Again there is ample heterogeneity ranging from a mean WTP of $0.21 to $3.40.[7] Also within person, the average standard deviation is $1.11 indicating that people have different valuations across the products.

In Figure 3 we plot purchase frequency as a function of WTP minus price. The dots represent the mean purchase frequency for individuals at that implied surplus (WTP minus price). Purchase frequency fits a smooth S-shaped curve. This would seem to be a success for the BDM as a method. As a benchmark, we plot the dotted step function which represents the actual predictions of the BDM. To optimists, the data fit surprisingly well. To pessimists, there is a large degree of error. Our interpretation leans more toward pessimism, noting that the average WTP of the goods in the sample is $2.03. Consider a consumer surplus of $|1|

[7] One possible explanation for this is that some people are considerably hungrier than others, or that their hunger plays a larger role in their calculation of WTP.

Figure 2: Distributions of WTP

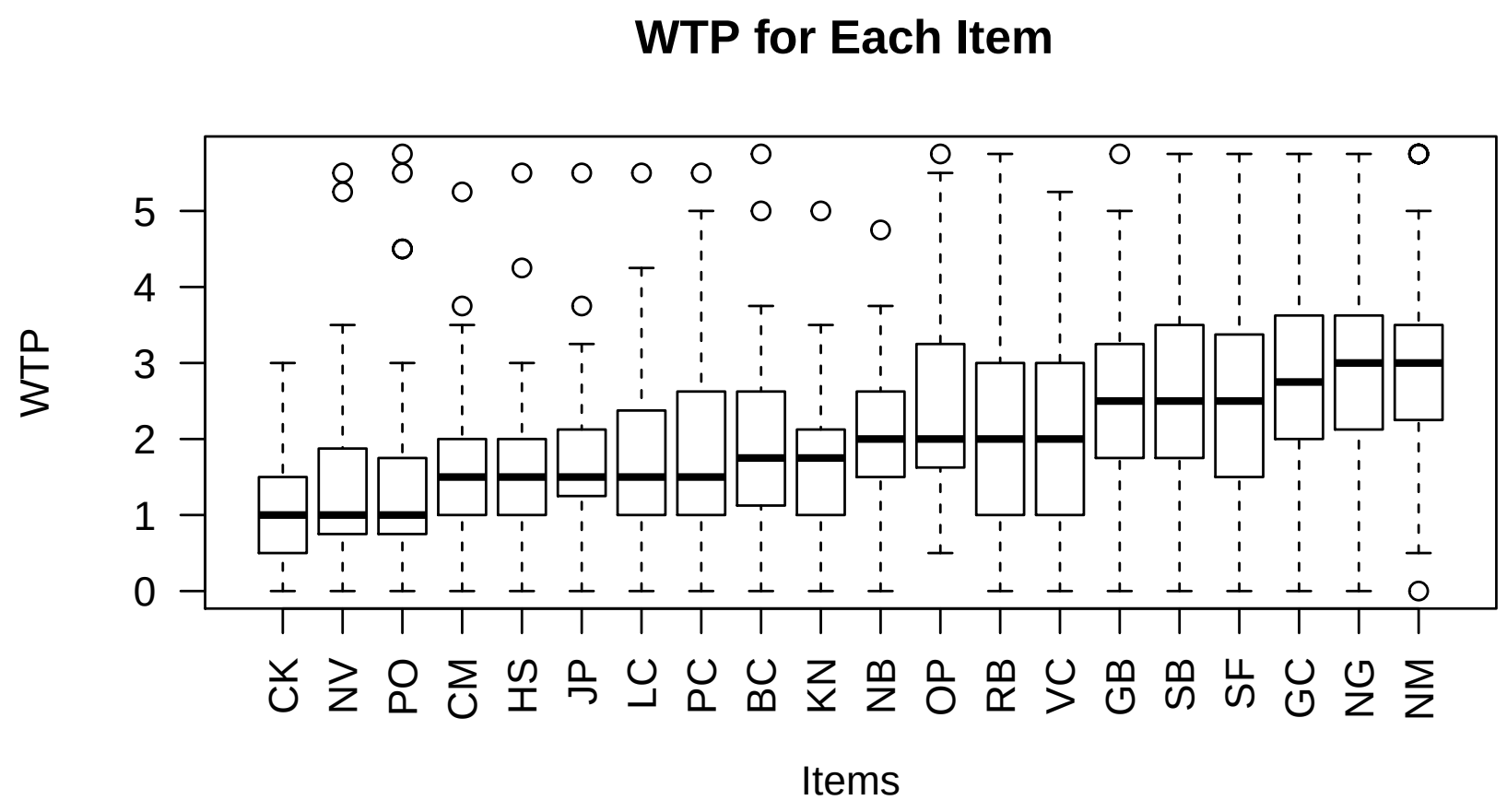

(a) WTP Distributions by Item

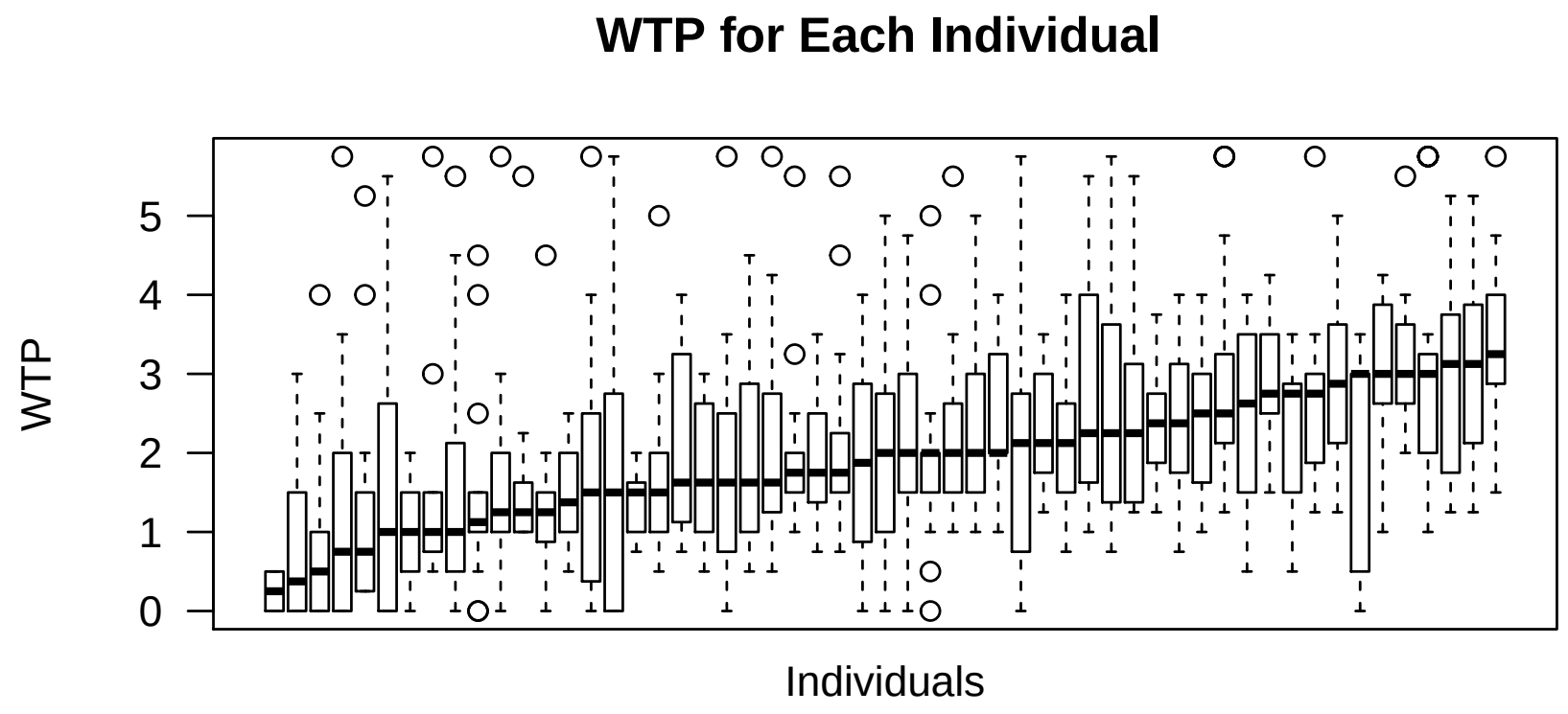

(b) WTP Distributions by Individual

Notes: All WTP are stated in US dollars. Food items and abbreviations are as follows. Bar Cliff Peanut Crunch BC; Chex Mix CM; Coke CK; Godiva Dark Chocolate GC; Green & Blacks Organic Chocolate GB; Hershey's Bar HS; Justin's Peanut Butter JP; KIND Nuts & Spices KN; Luna Choco Cupcake LC; Naked Green Machine NG; Naked Mango NM; Naturally Bare Banana NB; Nature Valley Crunchy NV; Organic Peeled Paradise OP; Pretzel Crisps Original PC; Pringles Original PO; Red Bull RB; Simply Balanced Blueberries SB; Starbucks Frappuccino SF; Vita Coco VC.

in Figure 3, which represents a roughly 50% change in the value of the average good. Even at such large surpluses relative to product value, purchases are still roughly 10 percentage points away from the BDM prediction.

Figure 3: Purchase Frequency as a Function of Surplus

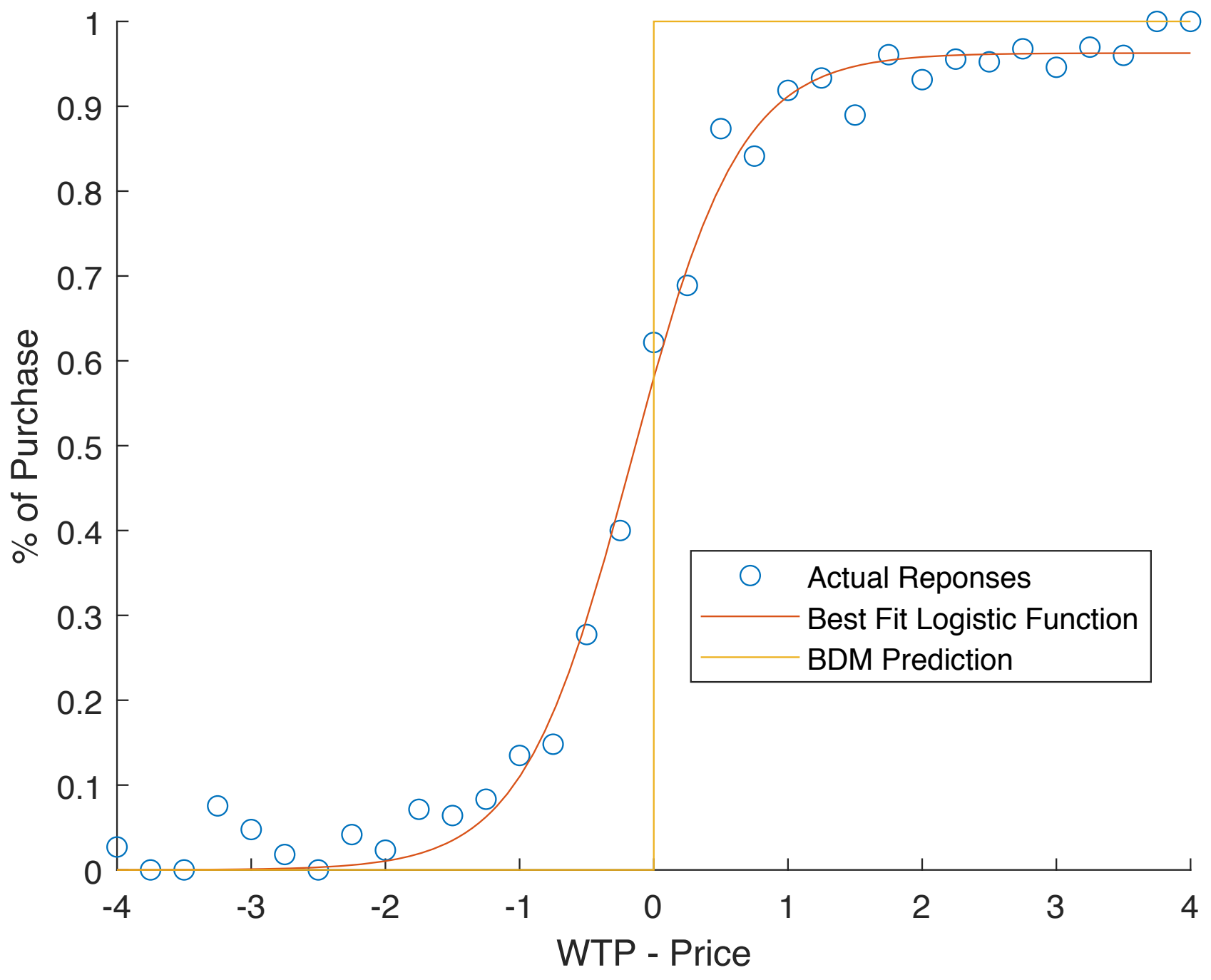

Notes: Each dot represents the probability of purchasing for a given bin with the specified consumer surplus (i.e., WTP - Price).

The second important weakness of the BDM prediction is a systematic bias. This is evident by noting that the probability of purchase at 0 surplus is 62%. This suggests a systematic understatement of people's reservation value. Intuitively, this may make sense if people use a bargaining heuristic when stating their WTP (Plott and Zeiler, 2005). If people view their stated WTP as what a person would verbalize in a bargaining context, they might always state a WTP below their true reservation price in order to obtain positive surplus conditional on obtaining the item. In addition, research has shown that the way the options and distribution of prices is presented in the BDM affects people's choices (Urbancic, 2011; Tymula et al., 2016). There may also be differences in the cognitive process underlying contingent choices versus non-contingent choices (Brandts and Charness, 2011). Another interpretation starts from the view of stochastic choice. People may not actually have static reservation prices, but instead reservation prices that fluctuate over time. Obtaining the good in the future at even a known price is then a risky lottery. If people are risk averse and/or loss averse this could deflate people's stated WTPs as people may be concerned that their tastes will change by the time they actually receive the item. A third interpretation is that as the experiment progresses, demand for the food items goes up. The BDM-Task always took place before the Buy-Task. People may have become hungrier over time or the constant exposure to pictures of food may have increased demand. While this is possible we do not observe robust evidence of this (see Appendix B for details).

The experiment was not designed to test between these mechanisms. We document the downward bias here and note that it is an important justification for finding alternative methods for measuring individual valuation (Miller et al., 2011). One approach to "debias" BDM choice predictions would be to use a model $buy_{ijt} = 1\{WTP_{ij} - p_{ijt} + x \geq 0\}$ instead of the standard BDM prediction of $buy_{ijt} = 1\{WTP_{ij} - p_{ijt} \geq 0\}$. One could then find the $x$ that minimizes MSE. The scale we employed to elicit WTP used intervals of 0.25, so this limits our level of granularity. We found that $x = 0$ performs better than $x = 0.25$, $x = 0.50$, or larger intervals. However, if we used smaller increments it is likely that a strictly positive $x$ would minimize MSE given the apparent a bias.

Most crucial for our purposes, visual inspection reveals that the logit curve representing the predicted values from a logit regression in Figure 3 fit the data better than the BDM prediction. In other words, it is clear that there are gains from SML. We present quantitative evidence in the next section.

# 6 Supervised Machine Learning

## 6.1 WTP Data

We now compare SML to the direct elicitation in predicting purchases on the Buy-Task. We use the same dataset, the output of the BDM-Task. We reserve 440 observations (10% of our sample) to compute out-of-sample MSE, as detailed in Section 4.3.

Consideration of the results begins with our benchmark models plotted in Figure 4(a). Like the rest of Figure 4, it displays the out-of-sample MSE as a function of the sample size in the training sample, progressing in intervals of 200, starting at 600. We benchmark performance against a model that predicts that any decision is "Buy" with probability $r$, where $r$ is the mean purchase frequency. This sets a lower bound for performance. We find that $r = 0.556$ and the MSE $= 0.247$ for the full sample. The second benchmark, BDM, has a MSE of 0.1473, well below the MSE of Prob(Buy), and constant across sample size. This is because there is virtually no fitting of the model. The single parameter is the probability that the person purchases the good when the price exactly equals the WTP. We use the population average in the training sample, and, due to the law of large numbers, this doesn't change much as our sample increases. The only other variation comes from the hold-out sample, and here the variance is reduced since we average over 50 random samples.

Figure 4: Performance of Supervised Machine Learning with WTP and 2AFC Data

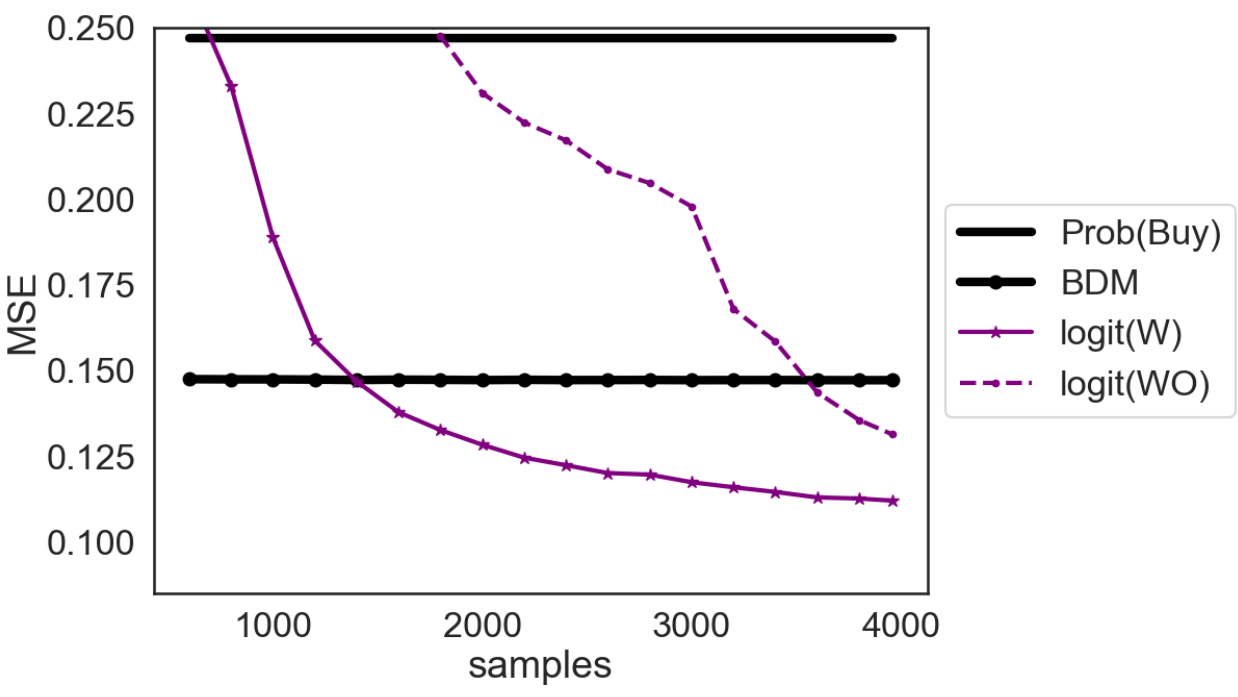

(a) Logit results with WTP data.

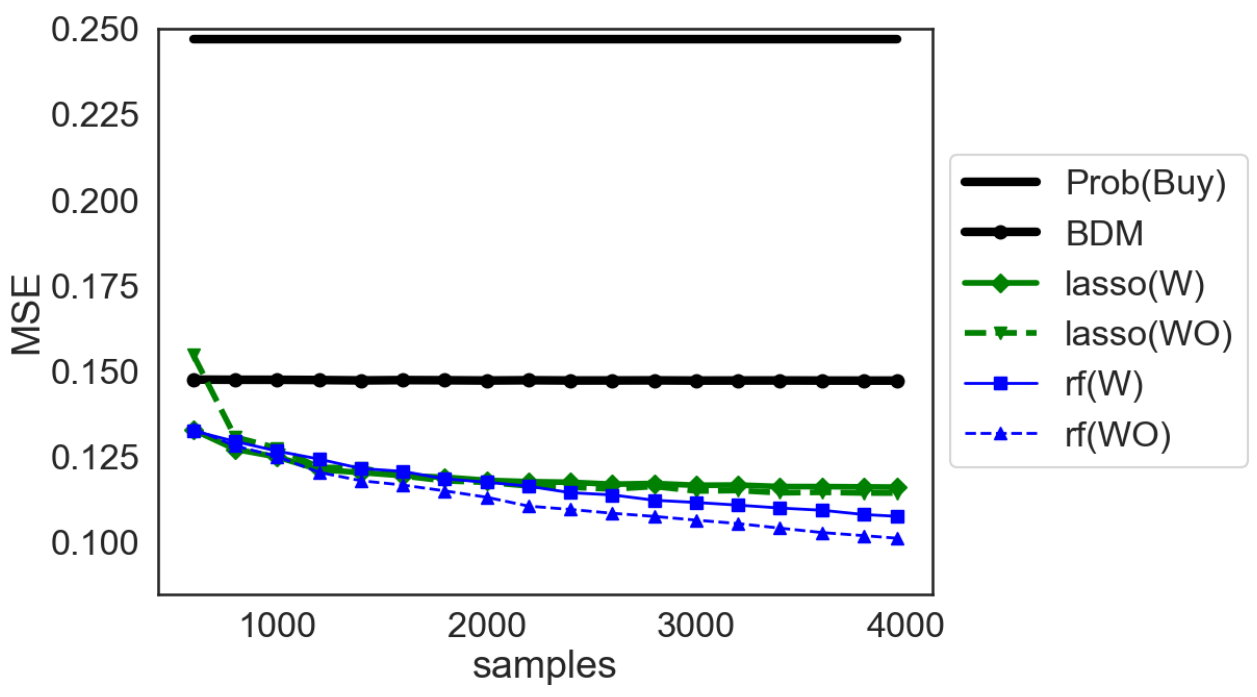

(b) Lasso and random forest results with WTP data.

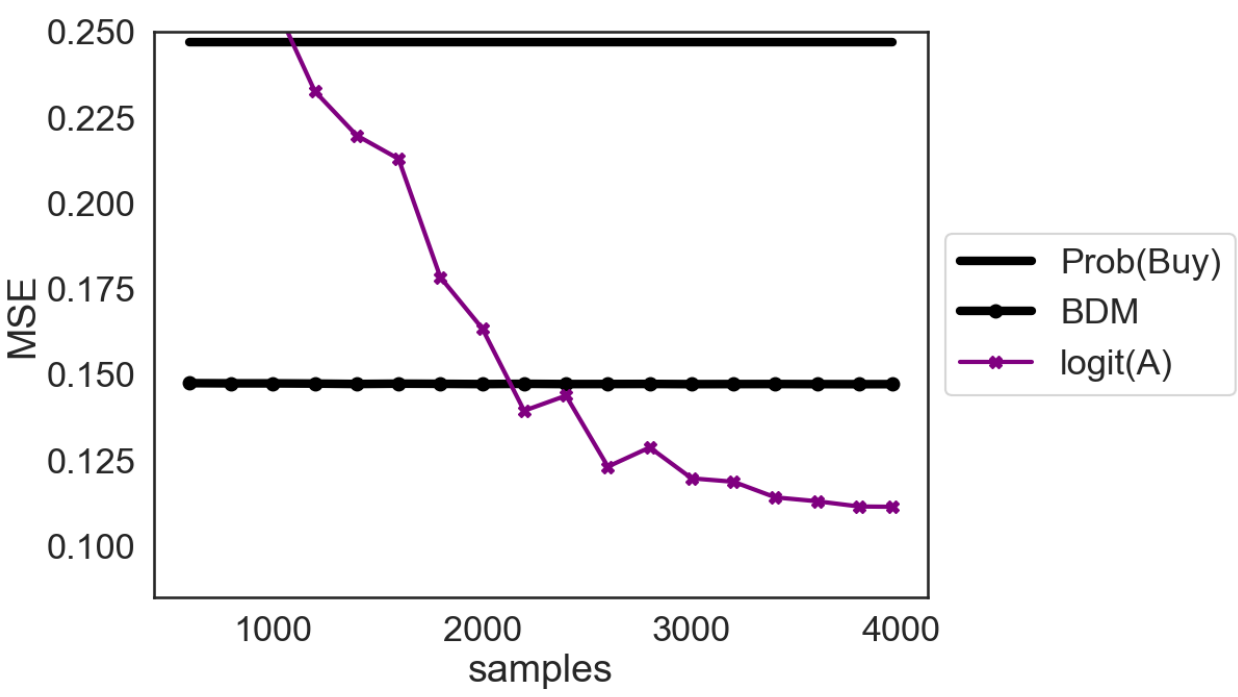

(c) Logit results with 2AFC data.

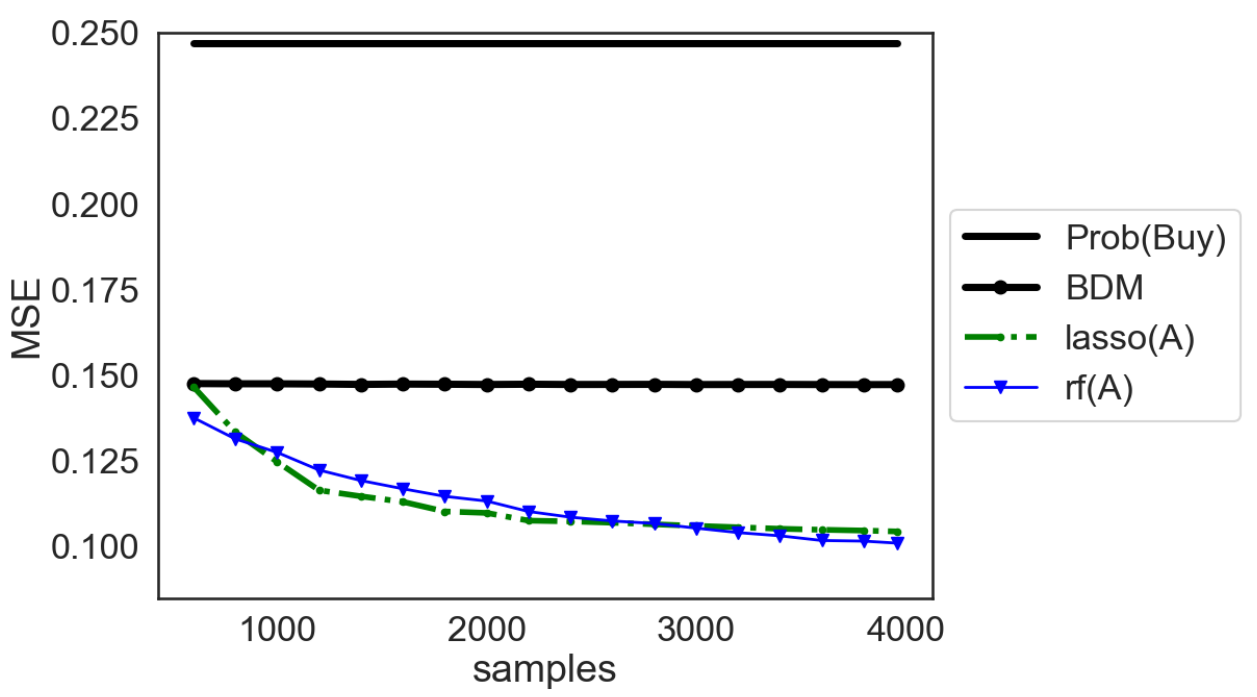

(d) Lasso and random forest results with 2AFC data.

Notes: Out-of-sample MSE estimated on 440 hold-out observations. Sample size increases from 600 to 3960 in intervals of 200. Since samples are random, we repeat the estimation 50 times and MSE is averaged.

Several clear trends emerge for the models using WTP data in Figures 4(a) and 4(b).[8]

[8]For ease of comparison, we use the same axes in all plots in Figure 4. A companion pair of plots for the logit models with extended axes is provided in the Appendix in Figure A2. We also want to note one limitation of the logit model. For training samples under 3000 observations, we frequently encountered rank-deficient matrices during model fitting. This translated to greater variability in out-of-sample predictions. We include the results for all samples to illustrate this limitation of logit for smaller training samples. This

First, at 1400 observations and higher, logit(W), our most basic statistical model, beats the BDM. The improvement continues as sample size increases, but most of the gains occur before 3000, ending with an MSE of 0.1121. We contrast this with our most sophisticated analysis of the same feature space, random forest. From the plot in Figure 4(b) it is clear that rf(W) outperforms the BDM from the smallest sample size, 600. It exhibits continual improvement throughout our entire training sample, up to 3960 observations, with a final MSE of 0.1077. This shows the considerable benefit of using high-dimensional data with high-dimensional methods.

We next decompose the gains by data and method. The logit(WO), shown in Figure 4(a), uses the full vector of WTP and begins with a performance of 0.467 MSE at 600 observations, doing even worse than Prob(Buy). The MSE declines rapidly but only beats the BDM at 3600 observations and achieves a MSE of 0.131 at the maximum sample, worse than its lower-dimension sister model, logit(W). This shows that using high-dimensional data with a low-dimensional method results in poor performance. As discussed previously, rf(W) outperforms the BDM at all sample sizes, and the same is true for rf(WO). Furthermore, rf(WO) outperforms rf(W) starting at a sample size of 800, with the gap growing until the full number of training observations is reached. The final MSE for rf(WO) is 0.1013, lower than that of lasso(WO). Likewise, rf(W) outperforms logit(W). This evidence shows that the gain in performance comes from using high-dimensional data with high-dimensional methods. Cross matching the data with the method produces worse results than the more basic low-dimension data, low-dimension method, logit(W). Visual inspection of the lasso(W) and lasso(WO) results show that lasso(WO) also outperforms lasso(W), starting at 1600 observations. The gains are more modest, though, with a final MSE of 0.1162 for lasso(W) and 0.1145 for lasso(WO).

## 6.2 2AFC Data

Can we predict purchases at specific price levels using the 2AFC data — data which provides an ordinal ranking of alternatives but contains no cardinal information about reservation values? The answer is yes. All of the statistical models beat the baseline BDM benchmark. For the logit(A) model, Figure 4(c) shows the MSE as a function of the sample size of the training sample. At roughly 2600 observations and higher, the model performs better than the BDM. For the high-dimensional methods, shown in Figure 4(d), superior performance comes almost immediately, as prior to 1000 observations all lasso and random forest models applies for both the WTP and 2AFC data.

are better.

Surprisingly, we see no gain from using RT data. All models using RT data perform worse initially than their sister models without RT. As the sample size grows, the models with RT approach or converge to the performance of their sister model without RT. With the full sample, sister models without and with RT have almost the same MSE for both lasso and random forest. More details, including a corresponding plot in Figure A5, are provided in the Appendix.

Both lasso and random forest with 2AFC data perform well, beating out logit(A). Inspecting the plot shows that the two high-dimensional models asymptotically approach different performance levels, suggesting more data is not a substitute for a better algorithm. For lasso, the MSE at the full sample is 0.1044, and for random forest it is 0.1010, compared to 0.1115 for logit(A).

Perhaps the most stark finding comes from comparing the results for 2AFC data to WTP data. The MSE (standard errors in the parentheses) for rf(A) at the full training sample, at 0.1010 (0.0010), is approximately the same as the MSE for rf(WO), 0.1013 (0.0012). For lasso, 2AFC data has a smaller MSE, at 0.1044 (0.0012), compared to 0.1145 (0.0012) for the full WTP dataset. Even a conservative interpretation would grant that the 2AFC data provides equivalent out-of-sample prediction to the WTP data for subsequent purchase behavior. The 2AFC data produces high-dimensional information on each subject's ranking of the food items. Cardinal information on the value of each item is entirely lacking in the raw data.

It is worth noting that prediction here is both within-subject and between-subject. A given individual's buy decisions may be split up into the training sample and the test sample. The question we are answering is whether one can predict the behavior of individuals with 2AFC data given their past purchase behavior. A different question is whether one can predict the purchase behavior of a set of individuals using only data from a different set of individuals. This latter exercise would be an entirely between-subject prediction.

We posit that the strong performance of the rf(A) model comes from high quality prediction within individuals since the 2AFC data represents ordinal-ranking data within person. We suspect that the between-subject performance with 2AFC data would do worse because there is no direct data on the relative valuations of goods between subjects. We return to this question in Section 7.4.

In summary, we find that high-dimensional data with a high-dimensional method offers the best predictions (i.e. rf(WO)). The WTP data and the 2AFC data are nearly equally predictive of buy decisions when one uses random forest (i.e. MSE of rf(WO) $\approx$ MSE of rf(A)). But much of the gain relative to the BDM benchmark is achieved through a simple logit (i.e. logit(W)).

# 7 Optimizing Data Use

## 7.1 Opening the Black Box

The random forest algorithm yields models that escape conventional interpretation, such as comparing regression coefficients. These models generate many regression trees and then average over these trees. In the original paper, Breiman (2001) detailed other statistics that can shed light inside the black box. We explore two such measures: the mean decrease in accuracy and the mean decrease in Gini index in the rf(WOA) model, the best performing model. The mean decrease in accuracy is the increase in prediction error from taking a feature and randomizing it without replacement across the observations. The Gini index is the total decrease in prediction error from splitting on the variable, averaged over all trees (James et al., 2015).

According to these metrics, the best predictors overall are price, polynomial expansions of price, the percentage of times the item was chosen in 2AFC trials, the rank of the item in terms of times chosen in 2AFC trials, WTP, and item fixed effects. In other words, the random forest algorithm leveraged distinct pieces of data from the Core, WTP, and 2AFC data. This suggests that the WTP data and 2AFC data might not be redundant. We address this question next.

## 7.2 Are WTP and 2AFC Data Redundant?

One way we can consider the possible redundancy of WTP and 2AFC data is by including both datasets in a statistical model. Figure 5 shows the results for all training sample sizes. For the random forest algorithm, presented in Figure 5(a), we include several feature space combinations for comparison, starting with the baseline Core model, rf(C). The model of interest is random forest on the union of the WTP and 2AFC features called rf(WOA), which yields better performance at all sample sizes. MSE decreases continuously across the sample with a minimum of 0.0890 (0.0009) when the full training sample is used. This MSE is less than the 0.1013 (0.0012) for rf(WO) and 0.1010 (0.0010) for rf(A). This shows that there is indeed some non-redundant information in the two data sets.

Figure 5: Performance of Supervised Machine Learning with All Combinations of Data

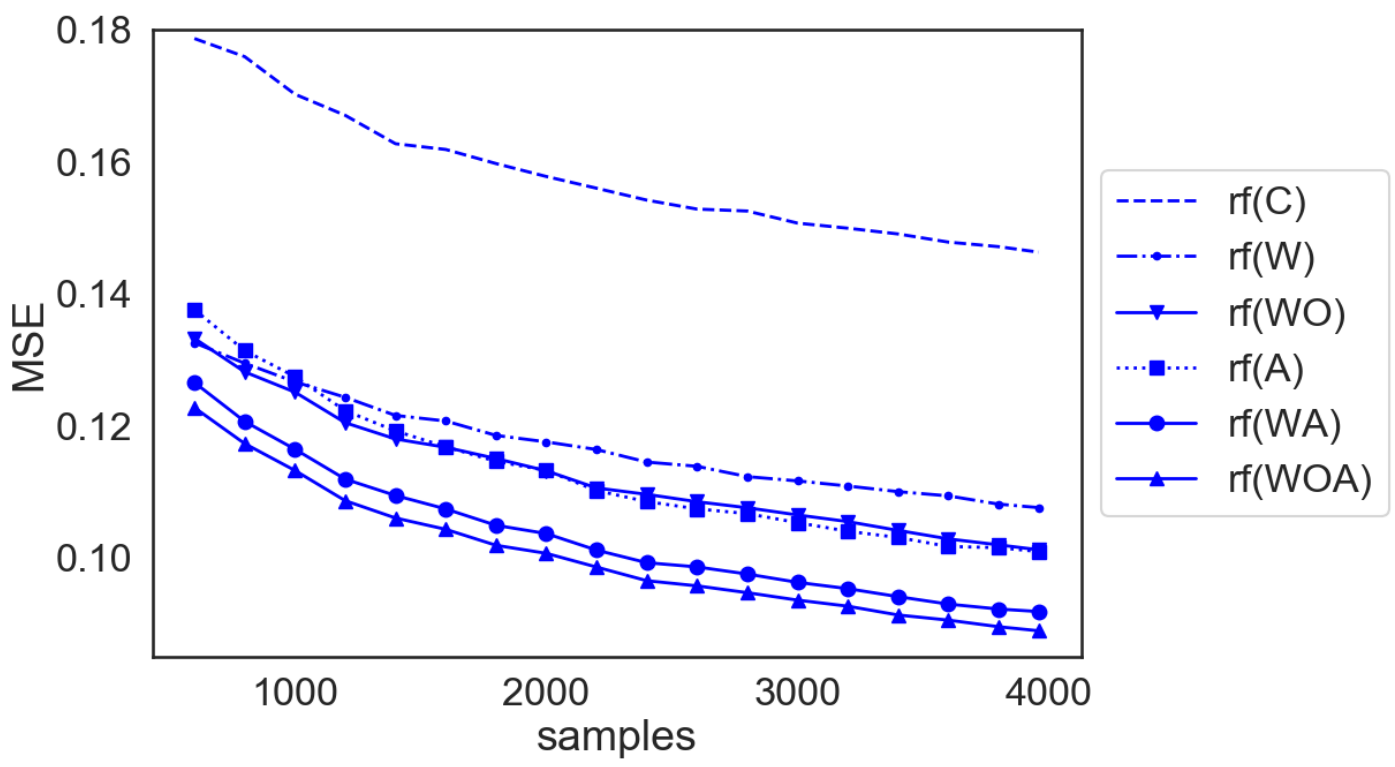

(a) Results for random forest algorithm.

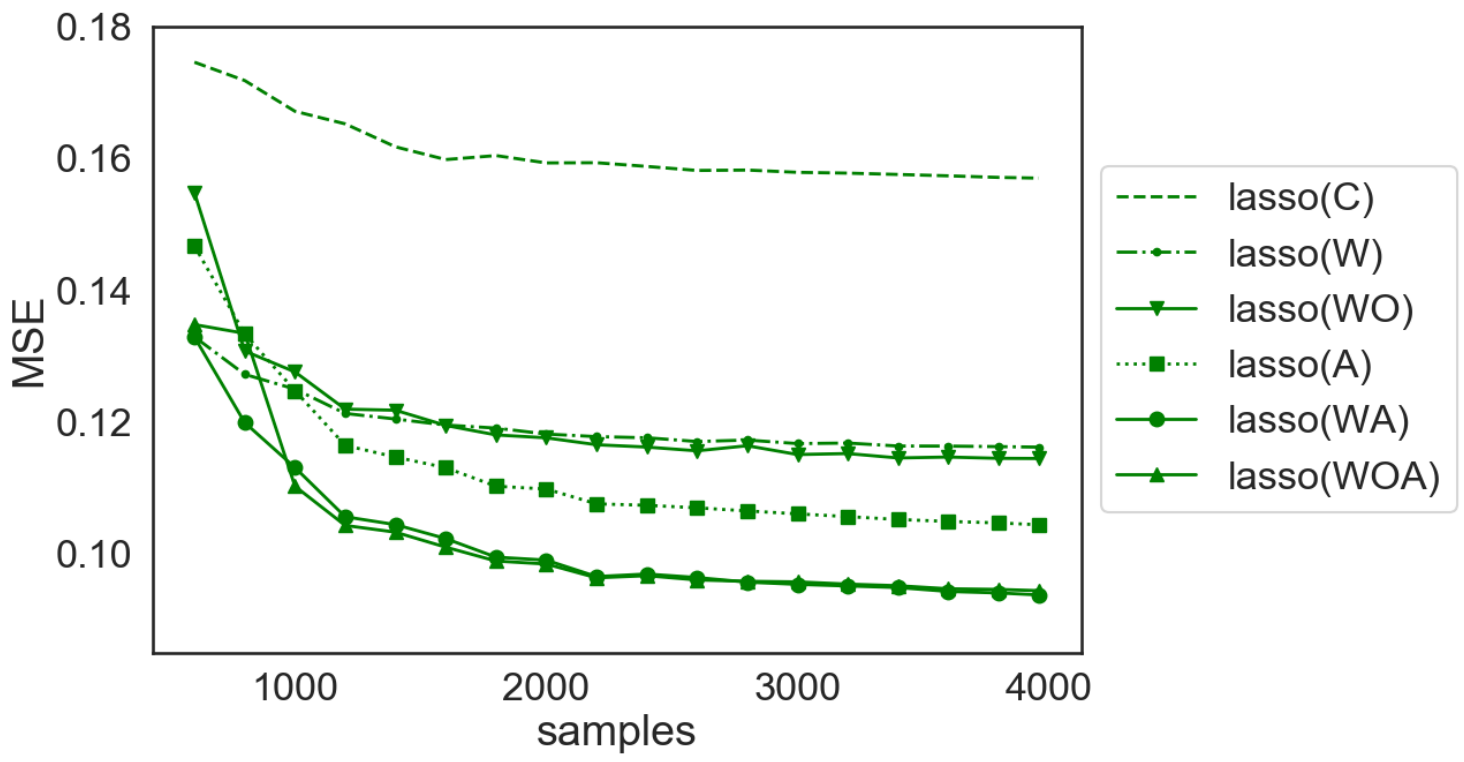

(b) Results for lasso algorithm.

Notes: Out-of-sample MSE estimated on 440 hold-out observations. Sample size increases from 600 to 3960 in intervals of 200. Since samples are random, we repeat the estimation 50 times and MSE is averaged.

Seeing the improvement from WO and A to the combined WOA, a practitioner might also be interested to know if a slightly smaller feature set could accomplish the same improved prediction. One option would be to simply focus on the WTP data for the item for which a practitioner wishes to predict purchase behavior. In other words, combine W with A feature spaces. From Figure 5(a) we can see that rf(WA) outperforms rf(W) and rf(A) at all sample sizes, but underperforms relative to rf(WOA).

For comparison, lasso results for the same combinations of features are shown in Figure 5(b). In general, all of the trends observed with random forest are also found in the lasso results. Clearly, the combination of WTP and 2AFC data improves predictions. The combination WOA has an out-of-sample MSE with the full training set of 0.0944 (0.0011), smaller than lasso(A) and lasso(WO). Interestingly, there is no gap between lasso(WOA) and lasso(WA). However, the MSE for lasso(WA) and lasso(WOA) are both greater than the respective MSE for rf(WA) and rf(WOA).

## 7.3 Where Does the BDM Get Reservation Value Wrong?

In Figure 6 we plot the MSE of the BDM as a function of the surplus as implied by WTP. Not surprisingly, most of the error comes from when the surplus is near zero. The interesting part is that MSE of the BDM hits a maximum at a surplus around -$0.25. This provides more evidence WTP from a BDM systematically understates reservation value. In other words, Figure 6 shows that WTP leads to more incorrect predictions of not buying (i.e., when price > WTP) than buying (i.e., when WTP > price).

We also plot the MSE of our best-performing SML model, rf(WOA). Random forest models rf(WO) and rf(A) are also plotted for comparison. As can be seen from Figure 6, the gain from the SML models is mostly in the surplus range of -$0.75 and $0.75. All models perform approximately the same outside this range, suggesting an irreducible unpredictability given the inherent stochasticity of choice data.

## 7.4 Between-Subject Analysis and Between-Item Analysis

The 2AFC data predicts remarkably well given its lack of cardinal information. Thus far, we have been predicting out-of-sample purchase decisions using both data within-person and between-person. Since 2AFC gives detailed ordinal information about a persons' demand within person, we suspect that performance might be driven by strong within-person prediction and slightly weaker between-person prediction. For example, there is nothing directly

Figure 6: Performance of BDM and Supervised Machine Learning as a Function of Surplus

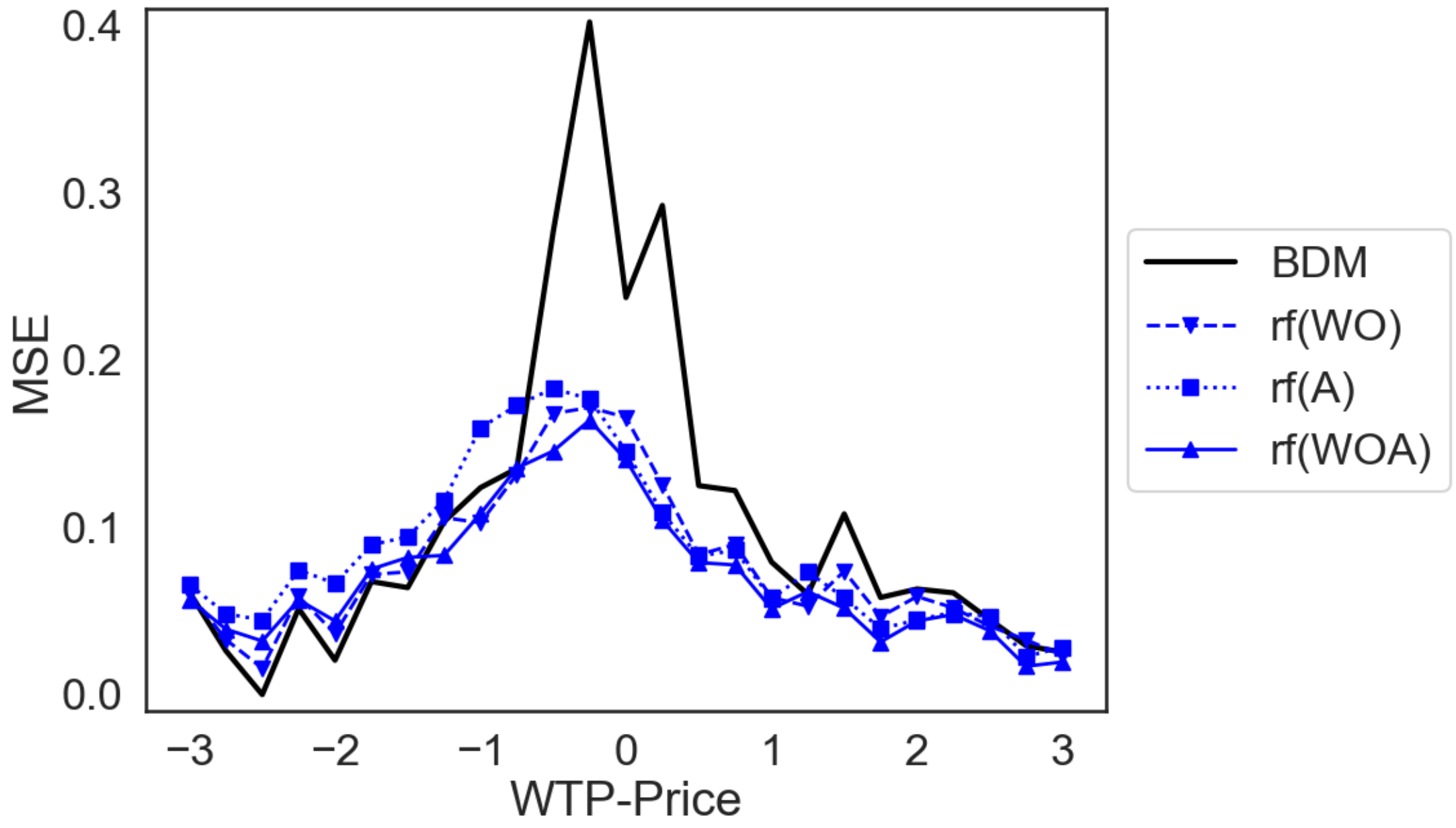

Notes: Out-of-sample MSE estimated on 440 hold-out observations. Since samples are random, we repeat the estimation 50 times and MSE is averaged.

in the 2AFC data that would indicate overall hunger or demand whereas this may be evident from heterogeneity in WTPs across subjects. On the other hand, there may be specific choice patterns that are indicative of overall demand. For example, very hungry people may exhibit a preference for high-calorie items and this might be correlated with high buy propensities on all items. Therefore it is worth testing how well models, in particular those employing 2AFC features, do when prediction is entirely between subject.

We conduct the between-subject analysis by randomly selecting 5 out of 55 subjects, with all their observations, to be the test sample. We estimate our models on the remaining 50 subjects, using cross-validation procedures similar to those outlined in Section 4.2. We repeat the random sampling 50 times and average our results. The MSE results are displayed in the second column of Figure 7 (standard errors in parentheses), using the same list of feature spaces as those displayed in Figure 5. All models perform worse than their respective baseline models (first column), but the drop in performance varies. As the out-of-sample performance for logit with the larger feature spaces was much poorer than for either lasso or random forest, this section will focus on lasso and random forest. A complete table with all results is provided in the Appendix in Table A2.

The emerging pattern is that between-subjects prediction increases MSE more for the 2AFC models than for the WTP models. Consider the lasso results first. The 2AFC model, lasso(A), performs better when using within&between-subject data, than the WTP model, lasso(WO), with respective full-sample MSE of 0.1044 (0.0012) and 0.1145 (0.0012). However, with only between-subject data, the lasso(A) MSE increases by 68.7% to 0.1762 (0.0044) while the lasso(WO) MSE increases only by 23.0% to 0.1408 (0.0035). We see the same pattern with the random forest models: rf(WO) and rf(A) perform almost equivalently at 0.1013 (0.0012) and 0.1010 (0.0010) MSE with within&between-subject data. In the between-subject analysis rf(WO) increases by 32.3% to 0.1340 (0.0027) and rf(A) increases by 42.1% to 0.1435 (0.0026). So, while the WTP data and 2AFC data perform about equally in between- and within-subject analysis, 2AFC data performs worse in between-subject analysis.

We conduct a similar between-item exercise in which we leave out 2 items and all their observations in the test sample. The only necessary change in feature spaces involves dropping the item fixed effects. We estimate our models on the remaining 18 items. As before, we repeat the random sampling 50 times and average our results. This would be akin to having data on an existing set of customers as well as their incentivized preferences for a novel good and then predicting whether they buy this novel good at various prices. These results are summarized in the third column in Figure 7. Again, the best performing model is

Figure 7: Test Sample — Holding Out Observations vs. Holding Out Subjects

MSE

| | Within&Between Subjects | Between-Subjects | Between-Items |
|---|---|---|---|
| lasso(W) | 0.1162 (0.0011) | 0.1301 (0.0022) | 0.1205 (0.0018) |
| lasso(WO) | 0.1145 (0.0012) | 0.1408 (0.0035) | 0.1193 (0.0017) |
| lasso(A) | 0.1044 (0.0012) | 0.1762 (0.0044) | 0.1160 (0.0017) |
| lasso(WA) | 0.0937 (0.0012) | 0.1331 (0.0032) | 0.0995 (0.0015) |
| lasso(WOA) | 0.0944 (0.011) | 0.1466 (0.0043) | 0.0988 (0.0014) |
| rf(W) | 0.1077 (0.0014) | 0.1573 (0.0034) | 0.1224 (0.0027) |
| rf(WO) | 0.1013 (0.0012) | 0.1340 (0.0027) | 0.1248 (0.0023) |
| rf(A) | 0.1010 (0.0010) | 0.1435 (0.0026) | 0.1138 (0.0015) |
| rf(WA) | 0.0919 (0.0009) | 0.1226 (0.0021) | 0.0986 (0.0013) |
| rf(WOA) | 0.0890 (0.0009) | 0.1187 (0.0020) | 0.0970 (0.0013) |

Notes: Summary table of out-of-sample MSE results for high-dimensional methods, using the full available sample of features. Standard error across the $N = 50$ runs is listed in parentheses.

rf(WOA). The models that use 2AFC data, lasso(A) and rf(A), perform better than their respective between-item analysis with the models that use WTP data, lasso(WO) and rf(WO), although the difference is slight with lasso. Overall, it appears that while WTP data is better for predicting between-subject, 2AFC data is slightly better for predicting between-item.

Visual inspection of the heatmap in Figure 7 reveals a clear pattern between columns. Whereas the average increase in MSE across all listed random forest models for the between-subjects analysis is 37.5%, the average increase in MSE for the between-item analysis is 13.2%. In other words, the increase in MSE is *smaller* for between-item than between-subject. A parallel result holds for lasso (40.2% for between-subjects, and 6.0% for between-items). In these cases, both algorithms face a greater challenge for prediction on new individuals compared to new items.

## 7.5 Which Elicitation to Use?

The results lead to a natural question: “When would a practitioner prefer to use WTP elicitations, 2AFC elicitations, or both?” WTP data predicts slightly better in our main analysis and much better in the between-subjects analysis. It is also important to take into consideration the costs of these elicitations. The average amount of time required for a 2AFC question is about 2 seconds and the average amount of time for a WTP question is 8 seconds. The time required to read the instructions for the 2AFC task was 1 minute 20 seconds, and for the BDM it was 4 minutes 30 seconds. There were a total of 190 2AFC questions and 20 WTP questions, which leads to an average length of 7 minutes 40 seconds for the 2AFC task, and 7 minutes 10 seconds for the BDM-Task. At the scale we used, the time costs are approximately equivalent. However, there are still some reasons to prefer the 2AFC-Task. If a researcher wants to keep the elicitation very short, the 2AFC-Task has a much shorter set of instructions and is more intuitive. Likewise, if a researcher has concerns that the population will struggle to understand the BDM rules, a 2AFC-Task may be preferred. If time is not an issue, a researcher may wish to conduct both elicitations as they may have different sets of strengths and weaknesses and, as several of our results have demonstrated, the information contained in the two is not redundant.

Perhaps there are other elicitations that generate even more predictive data. We speculate that there may be considerable gains in prediction from other elicitation methods and by the use of non-choice survey data (Netzer et al., 2008). Bernheim et al. (2013) show that this can be a productive avenue when trying to predict aggregate behavior.

# 8 Implications

## 8.1 Revenue-Maximizing Prices

Ultimately, the value of better prediction comes in the form of better action. There are many reasons why one may want to uncover a person's valuation for goods. Practitioners may wish to measure aggregate demand curves and heterogeneity of demand for a wide range of purposes, not limited to profit maximization, optimal taxation, and to measure the effects of an event or policy change. Mechanisms that include SML could potentially allocate resources more efficiently given that SML more accurately uncovers latent valuations. Perhaps the most obvious application would be to use the results of direct elicitation and SML for revenue maximization.

The BDM has an immediate implied revenue-maximizing price — the reported WTP. We can use our predictive model to estimate a revenue-maximizing price by simply maximizing $p \times \Pr(buy|p)$. To control for data quality, in both the BDM prediction and SML prediction we use the same WTP data for both. We therefore use the top-performing algorithm with this data, the rf(WO) model. Figure 8(a) displays a scatter plot of computed revenue-maximizing prices as a function of WTP, at the person-item level.

The revenue-maximizing prices are highly correlated with WTP ($\rho = 0.7378$). Nonetheless there is an obvious discrepancy. The average difference between WTP and the estimated optimal price is -\$0.2413 and the mean WTP is \$2.03, leading to an average difference of -11.9%. This difference again represents a systematic downward bias in WTP, as evidenced by more points being above the 45-degree line in Figure 8(a). It may be tempting to simply correct the BDM predictions by adding the average bias to the WTP to obtain a WTP$'$, but this would not lead to the correct conclusions as the price elasticities of the goods may vary. Indeed, the scatter plot in Figure 8(a) shows that this bias is not evenly distributed.

Looking at average absolute difference reveals the scale for potential improvement: the average absolute difference across all person-item pairs is \$0.5985, representing a 29.4% difference in the pricing of the product. Given that the SML model has substantially lower MSE, it is likely that the earnings of the rf(WO) would be considerably higher. We estimate the increased revenue from rf(WO) in the next section.

Figure 8: Revenue-Maximizing Prices

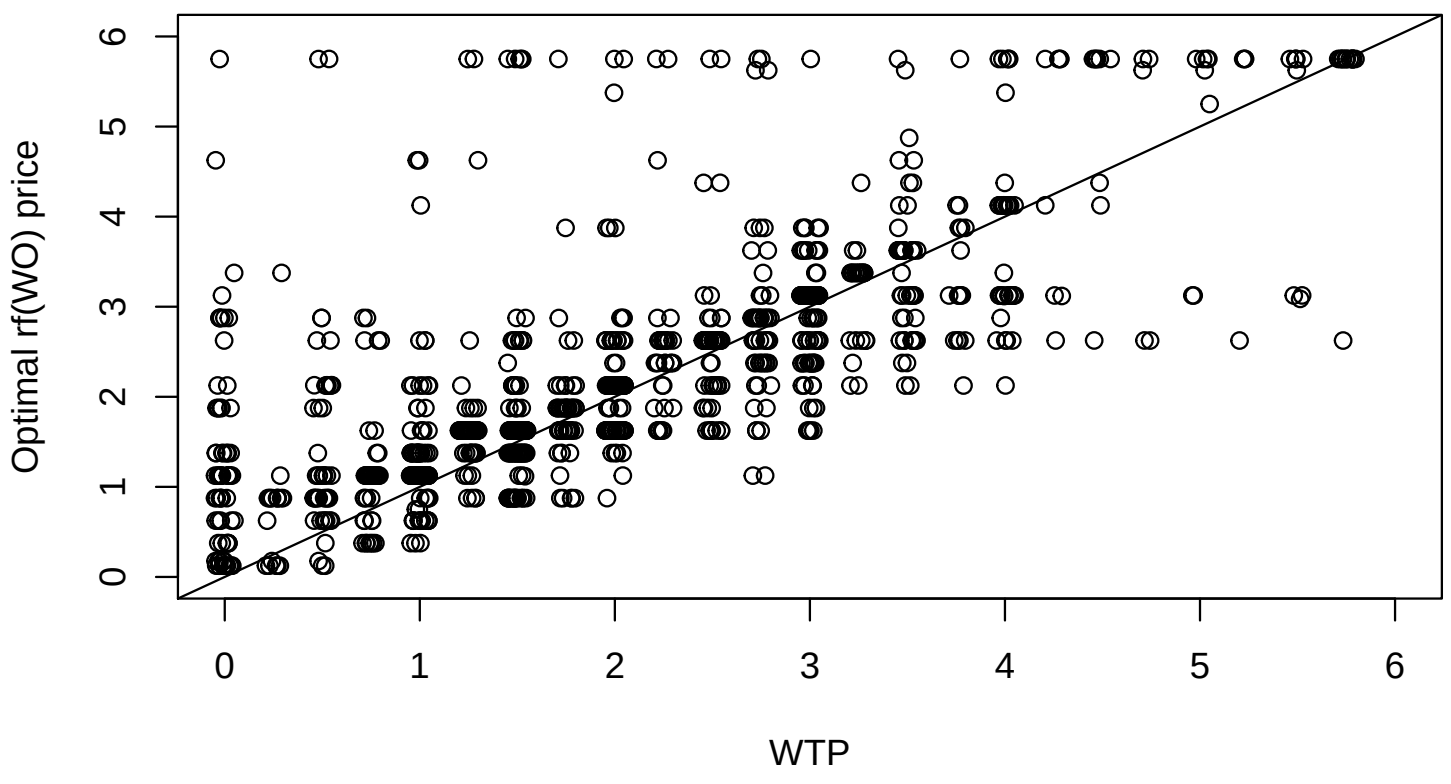

(a) Comparing WTP to the Revenue-Maximizing Price under rf(WO).

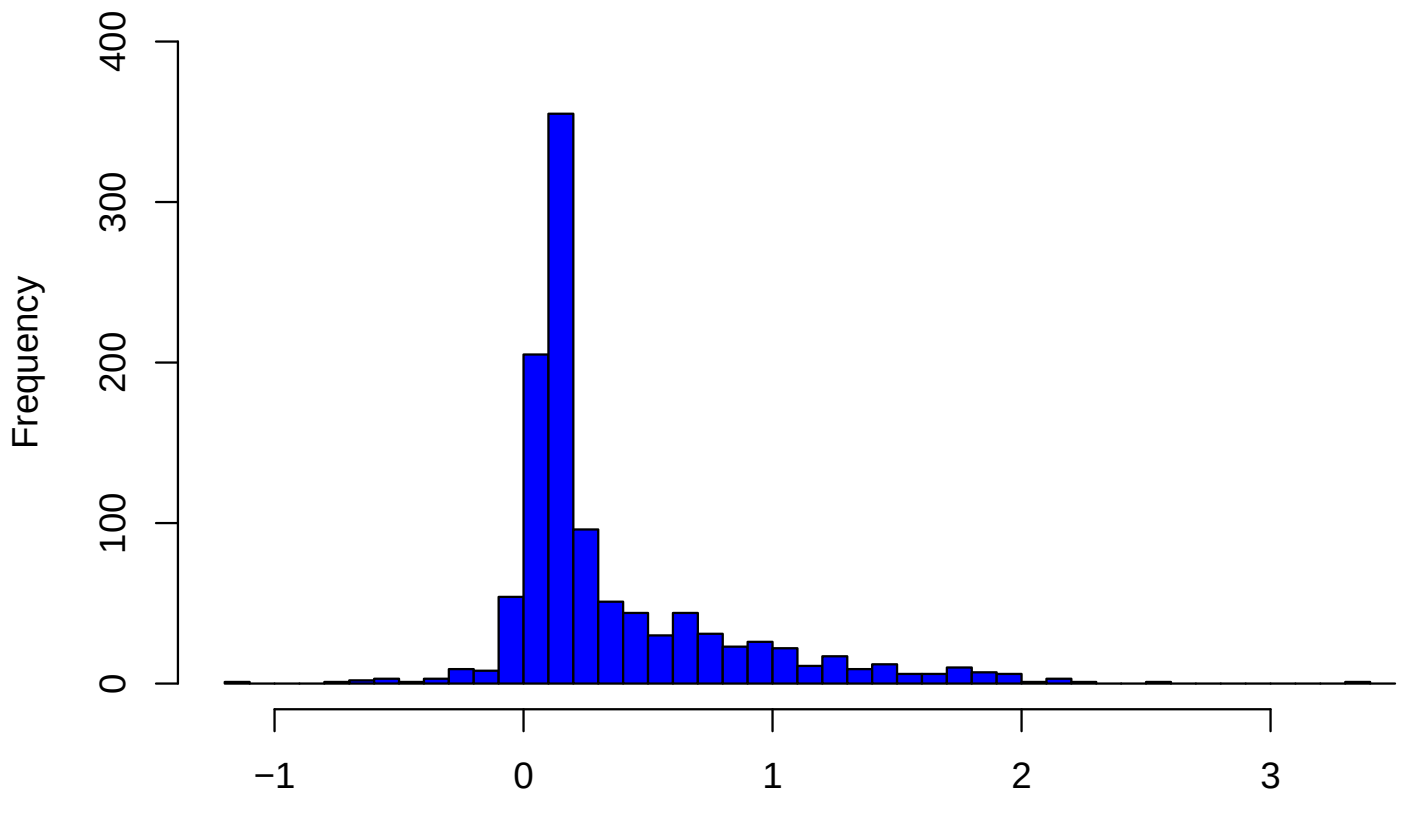

(b) Difference in Expected Revenue for Each Person-Item: Expected Revenue implied by rf(WO) minus expected revenue implied by WTP.

## 8.2 Estimated Revenue

In order to compare revenue under the revenue-maximizing prices we need to make some assumptions about the true probability of purchase as a function of price. Ideally, we want each individual's item-specific stochastic demand curves. In our setting, this cannot be perfectly observed, only estimated. We use rf(WOA), our best performing model, to estimate these stochastic demand curves. This model uses random forest on all of the 2AFC and WTP data. This model yields 1,100 individual$\times$item stochastic-demand curves. We then compare the expected revenue of setting the price equal to the person's WTP for an item, to the expected revenue of setting the price equal to the revenue-maximizing price under the rf(WO) model.

The average expected revenue of setting the price to the WTP is \$1.262 and the average expected revenue of setting the price equal to the revenue-maximizing price under the rf(WO) model is \$1.610. Thus, setting the price according to random forest yields an estimated increase in revenue by 27.6% over the raw WTP values without the use of any additional data. Figure 8(b) shows a histogram of the gain in revenue by using random forest across the 1,100 individual-item pairs. It is evident that the distribution's mass is predominantly positive (92.5% $> 0$). As a benchmark, the average gross margin across all sectors in the US is approximately 37% as of January 2019 (Damodaran, 2019). Thus the improvement from using SML over the BDM price is substantial.

# 9 Discussion

Direct elicitation is appreciated for the way it tightly connects theory to data. SML approaches are often promoted or disparaged, depending on the audience and context, for being "theory free." However, it is not our position that theory and SML are in conlfict. In fact, our view is that these two approaches are highly complementary. Direct elicitation, because it incentivizes behavior, and because it tightly links that behavior to the latent object of interest, may produce data that is highly predictive. When married to SML, direct elicitation data can be debiased, and neglected patterns in the data could be used in their own right. For example, we found that the WTP values of the 19 other goods were predictive of buying a given good.

Applied more broadly, SML could potentially be used to identify specific aspects of purchase decisions (like in our Buy-Task environment) that are not captured by WTP elicitation. For example, nutritional information or branding might play a larger role in purchase de-cisions as opposed to direct elicitation of reservation value. So long as the computational burden is manageable, our modular feature space method could be expanded to identify the most predictive kinds of data. If highly predictive features are not well explained by theory and/or not captured by WTP elicitation, a gap in theory has been identified.

One potential criticism of our SML approach is that we did not generate a reservation value but instead generated a stochastic demand curve. If one's purpose is to construct a reservation value for each individual-item then our exercise is incomplete. This is a reasonable objective if one is testing a theory that makes predictions about reservation values, or if stochastic demand curves are too unwieldy objects for comparison. Our SML approach can deliver this with one additional step: by constructing a mapping from stochastic demand curves to reservation values. For example, one can define the reservation value to be the price at which the person is exactly 50% likely to buy the good.

# 10 Conclusion

In this paper, we demonstrate that SML can be an effective alternative for recovering individual-level latent preferences. SML offers sizable advantages over direct elicitation. First and foremost, we show that using SML facilitates the combination of distinct kinds of preference-elicitation data to improve predictions of consumer demand. Nested within this contribution is the result that SML can enhance BDM predictions of purchase decisions. Second, we demonstrate that data generated from direct elicitation is not necessary for prediction of purchase decisions. Specifically, we indirectly elicit demand using 2AFC data, which is fast, easily comprehended by subjects, and predicts better than the BDM.

Our results also show that more data cannot substitute for a superior statistical algorithm. With our data, for instance, random forest does asymptotically better than logit, while logit does asymptotically better than the BDM. Indeed, statistical methods are not substitutes for each other, and neither are data types. Conversely, we find that elicitation (i.e., WTP) and choice (i.e., 2AFC) data are complementary, and can be combined to improve predictions.

The basic approach presented in this paper can be extended to estimating other individual-latent preference parameters such as risk preferences, time preferences, or option attributes of importance that are not directly observable to the researcher. We think that SML may be especially useful in capturing unobserved consumer heterogeneity given the difficulty consumers will have in directly stating such differences within most traditional elicitations.

The finding that SML models predict better than direct elicitations could have useful implications for theory. For example, the systematic bias in the BDM prediction suggests that there are other cognitive processes at play during the BDM that are not at play in everyday purchase decisions. This suggests that there may be more predictive theories waiting to be discovered, that are also be more parsimonious than SML models. Future work that could explain this gap may lead to improved theories of choice and innovative hypothesis generation.

# Online Appendix
## Supervised Machine Learning for Eliciting Individual Demand

# A Additional Information

## A.1 Food Items

Table A1: Food Items

| | Food Item | Abbrv. |
|---|---|---|
| 1 | Cliff Barr Peanut Crunch | BC |
| 2 | Chex Mix | CM |
| 3 | Coke | CK |
| 4 | Godiva Dark Chocolate | GC |
| 5 | Green & Blacks Organic Chocolate | GB |
| 6 | Hershey's Chocolate | HS |
| 7 | Justin's Peanut Butter Cup | JP |
| 8 | KIND Nuts & Spices | KN |
| 9 | Luna Choco Cupcake | LC |
| 10 | Naked Green Machine | NG |
| 11 | Naked Mango | NM |
| 12 | Naturally Bare Banana | NB |
| 13 | Nature Valley Crunchy | NV |
| 14 | Organic Peeled Paradise | OP |
| 15 | Pretzel Crisps Original | PC |
| 16 | Pringles Original | PO |
| 17 | Red Bull | RB |
| 18 | Simply Balanced Blueberries | SB |
| 19 | Starbuck's Frappuccino | SF |
| 20 | Vita Coco | VC |

## A.2 Features

Features used are presented in groups below. Numbers in parentheses reflect the number of features for each description.

### Core Features

Total: 149

- Price variables of the item being sold (3): $price_{ij}, price_{ij}^2, price_{ij}^3$
- Item indicators (19): $Item_{i2}, \ldots, Item_{i20}$
- Subject indicators (54): $Subject_2, \ldots, Subject_{55}$
- Interaction between Price of the item being sold and item indicators (19): $price_{ij} \times Item_{i2}, \ldots, price_{ij} \times Item_{i20}$
- Interaction between Price of the item being sold and subject indicators (54): $price_{ij} \times Subject_2, \ldots, price_{ij} \times Subject_{55}$

### Features Using WTP Data

Total: 76 Features

- WTP polynomial of the item being sold (3): $WTP_{ij}, WTP_{ij}^2, WTP_{ij}^3$
- Interaction between WTP of the item being sold and item indicators (19): $WTP_{ij} \times Item_{i2}, \ldots, WTP_{ij} \times Item_{i20}$
- Interaction between WTP of the item being sold and subject indicators (54): $WTP_{ij} \times Subject_2, \ldots, WTP_{ij} \times Subject_{55}$

### Features Using OtherWTP Data

Total: 400 Features

- WTP on all the items (20): $WTP_{i1}, \ldots, WTP_{i20}$
- Interaction between $WTP$ of each of the 20 items and each of the item indicators ($19 \times 20 = 380$): $WTP_{i1} \times Item_{i2}, \ldots, WTP_{i1} \times Item_{i20}, \ldots, WTP_{i20} \times Item_{i2}, \ldots, WTP_{i20} \times Item_{i20}$

## Features Using 2AFC Data

Total: 211 Features

- Whether the good in question was chosen over other items (20): $Choice_{ij1}, \ldots, Choice_{ij20}$
- Fraction of time that the item was chosen over other items (20): $Fraction_{i1}, \ldots, Fraction_{i20}$
- Fraction of time that the item $j$ being sold was chosen over other items (1): $Fraction_{ij}$
- Interactions between Fraction of time that the item in question was chosen over other items and item indicators (19): $Fraction_{ij} \times Item_{i2}, \ldots, Fraction_{ij} \times Item_{i20}$
- Interactions between Fraction of time that the item in question was chosen over other items and subject indicators (54): $Fraction_{ij} \times Subject_2, \ldots, Fraction_{ij} \times Subject_{55}$
- Standard deviation for subject $i$'s 20 $Fraction_{ij}$ variables (and polynomial expansion) (3): $stdFraction_i, stdFraction_i^2, stdFraction_i^3$
- Rank of each item by Fraction (20): $Rank_{i1}, \ldots, Rank_{i20}$
- Rank of the item in question, as determined by Fraction (1): $Rank_{ij}$
- Interactions between Rank of the item in question and item indicators (19): $Rank_{ij} \times Item_{i2}, \ldots, Rank_{ij} \times Item_{i20}$
- Interactions between Rank of the item in question and subject indicators (54): $Rank_{ij} \times Subject_2, \ldots, Rank_{ij} \times Subject_{55}$

## Features Using 2AFC RT Data

Total: 87 Features

- Response time (RT) of the item being sold $j$ against other items and polynomial expansion (40): $RT_{ij1}, \ldots, RT_{ij20}$, $RT_{ij1}^2, \ldots, RT_{ij20}^2$
- Interactions between response time and choice variables (20): $RT_{ij1} \times Choice_{ij1}, \ldots, RT_{ij20} \times Choice_{ij20}$
- Interactions between squared response time and choice variables (20): $RT_{ij1}^2 \times Choice_{ij1}, \ldots, RT_{ij20}^2 \times Choice_{ij20}$
- Mean and standard deviation of RTs within person and polynomial expansion (6): $meanRT_i, meanRT_i^2, meanRT_i^3, sdRT_i, sdRT_i^2, sdRT_i^3$
- For each item sold $j$ (1): $\sum_k \left[\frac{\max RT_i - RT_{ijk}}{\max RT_i} \times (2 \times Choice_{ijk} - 1)\right]$, where $k$ is all other items (i.e., pairwise choices between item $j$ and item $k$)

# B Additional Results

This section presents additional results to complement those included in the main text.

## B.1 Alternative Basic Comparison

Another natural starting point for the logit in Equation 1 in Section 4.1 would be to restrict $\beta_1 = -\beta_2$, $\zeta_k = -\delta_k$, and $\kappa_k = -\iota_k$ implying that the buy probability is increasing in consumer surplus, yielding:

$$\begin{aligned} buy_{ijt} =& \beta_0 + \beta_1(WTP_{ij} - p_{ijt}) + \sum_{k=2}^{20} 1\{k = j\} \cdot (\gamma_k + \delta_k(WTP_{ij} - p_{ijt})) \\ &+ \sum_{k=2}^{55} 1\{k = i\} \cdot (\eta_k + \kappa_k(WTP_{ij} - p_{ijt})) + \epsilon_{ijt} \end{aligned}$$

We tried this as well. The performance of this model is very similar to the model in Equation (1) in the main text. For simplicity, we present the results of that model only given its greater flexibility.

## B.2 Test of Increasing Demand Over Time

One possible concern is that people may have become hungrier over time or the constant exposure to pictures of food may have increased demand during the experiment. We can measure WTP within the BDM-Task and probability of purchase in the Buy-Task as a function of trial number (we do not observe demand in the 2AFC task). In the BDM-Task, using OLS, the effect of trial number is positive with a coefficient of 0.0095 and significant at p=0.08. This represents a 19% increase in WTP from the first trial to the 20th trial, which provides some weak evidence of increasing demand over time. In contrast, in the Buy-Task, using logit, the effect of trial number has a coefficient of -0.0024 and is significant at p=0.07. This represents an 18% decrease in the probability of a purchase between the first trial and 80th trial. Given the inconsistent pattern, there does not seem to be a systematic effect of exposure on demand.

## B.3 Summary of Additional Tables and Figures

Figure 3 in the main text provides a plot of purchase frequency as a function of WTP minus prices, across all items. Here, Figure A1 shows the same data, broken down item by item.

Table A2 presents all full-sample results for logit, lasso, and random forest, a subset of which are presented in the main text in Figure 7.

Figure A2 parallels Figure 4 in the main text for the logit models, but provides the full range of MSE performance for all training sample sizes.

Figures A3 and A4 present prediction performance of SML using area under the curve (AUC) instead of MSE.

Figure A5 presents out-of-sample MSE for lasso and random forest models that include RT data from the 2AFC-Task, as well as corresponding models without the RT features included.

Figure A1: Purchase Frequency as a Function of Surplus

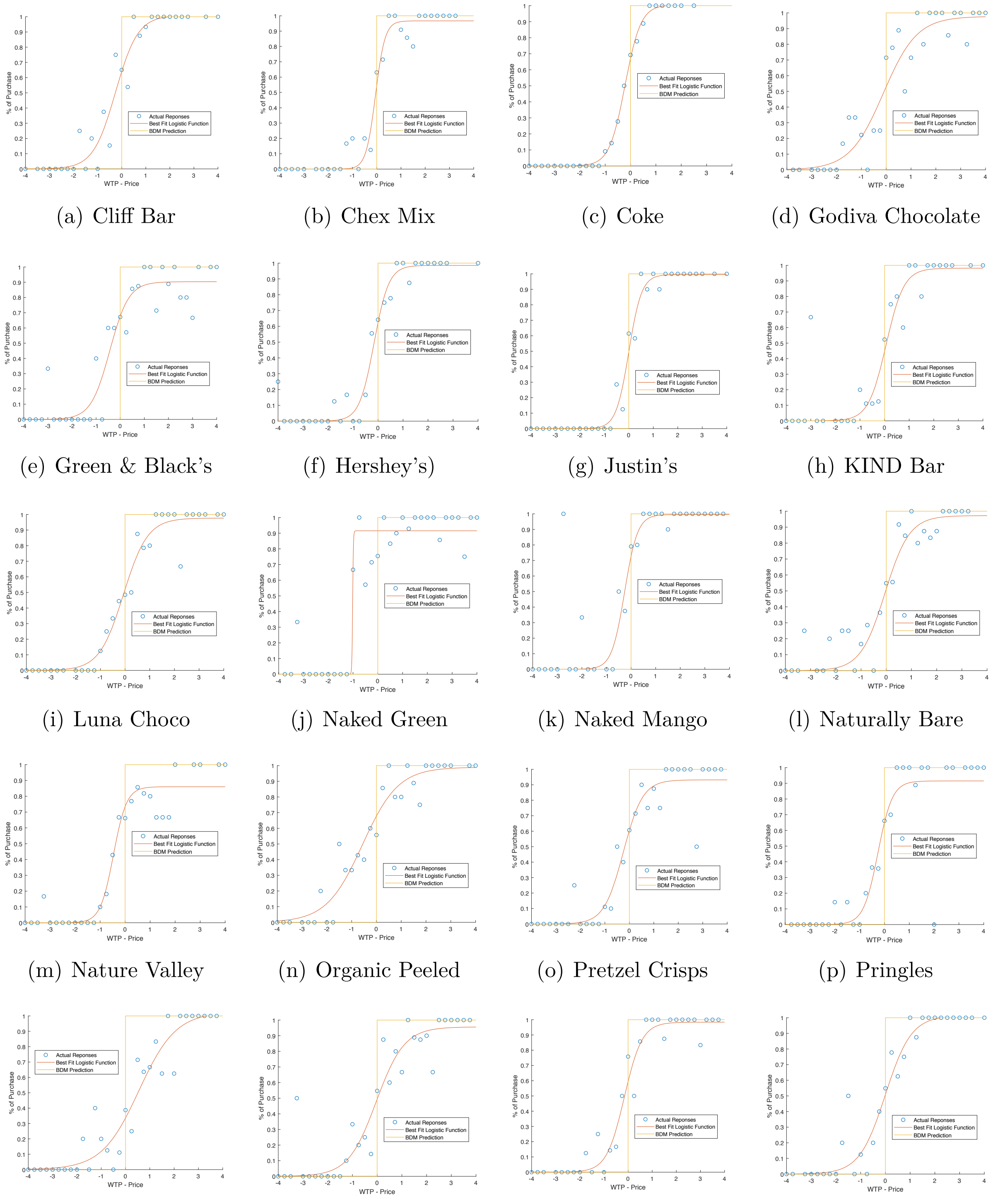

(a) Cliff Bar (b) Chex Mix (c) Coke (d) Godiva Chocolate

(e) Green & Black's (f) Hershey's) (g) Justin's (h) KIND Bar

(i) Luna Choco (j) Naked Green (k) Naked Mango (l) Naturally Bare

(m) Nature Valley (n) Organic Peeled (o) Pretzel Crisps (p) Pringles

(q) Red Bull (r) Simply Blueberries (s) Frappucino (t) Vita Coco

Notes: Each dot represents the probability of purchasing for a given bin with the specified consumer surplus. Items are also listed in Table A1.

Figure A2: Performance of Supervised Machine Learning with WTP and 2AFC Data

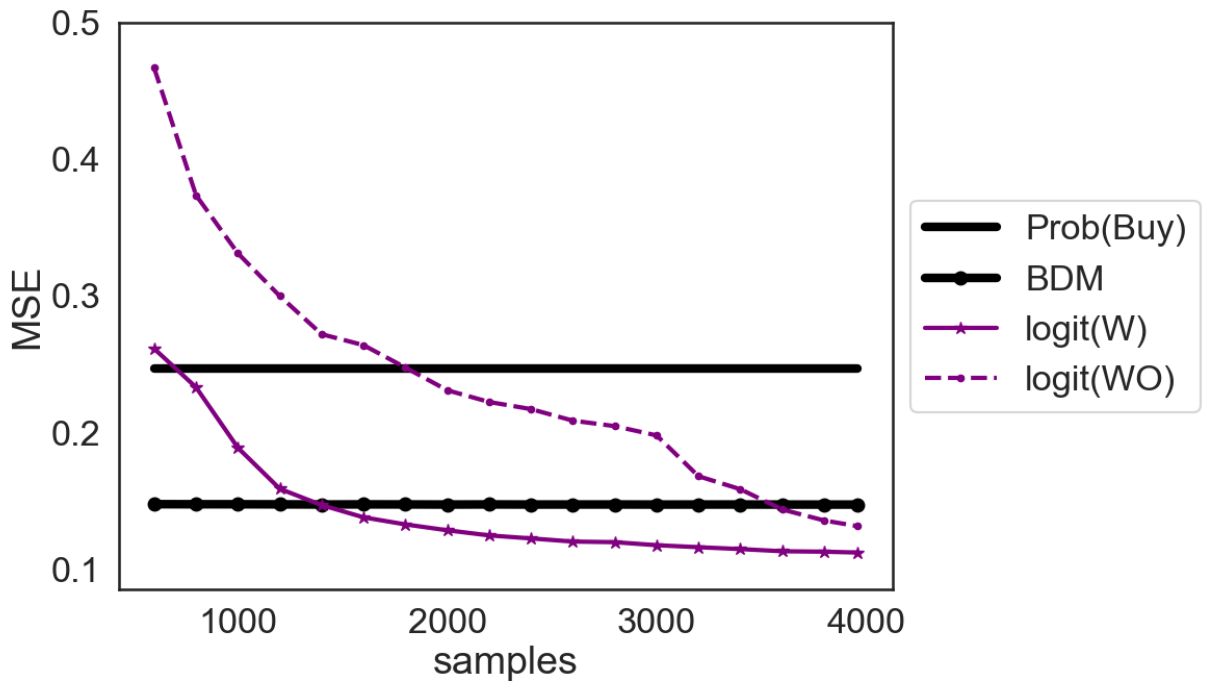

(a) Logit results with WTP data.

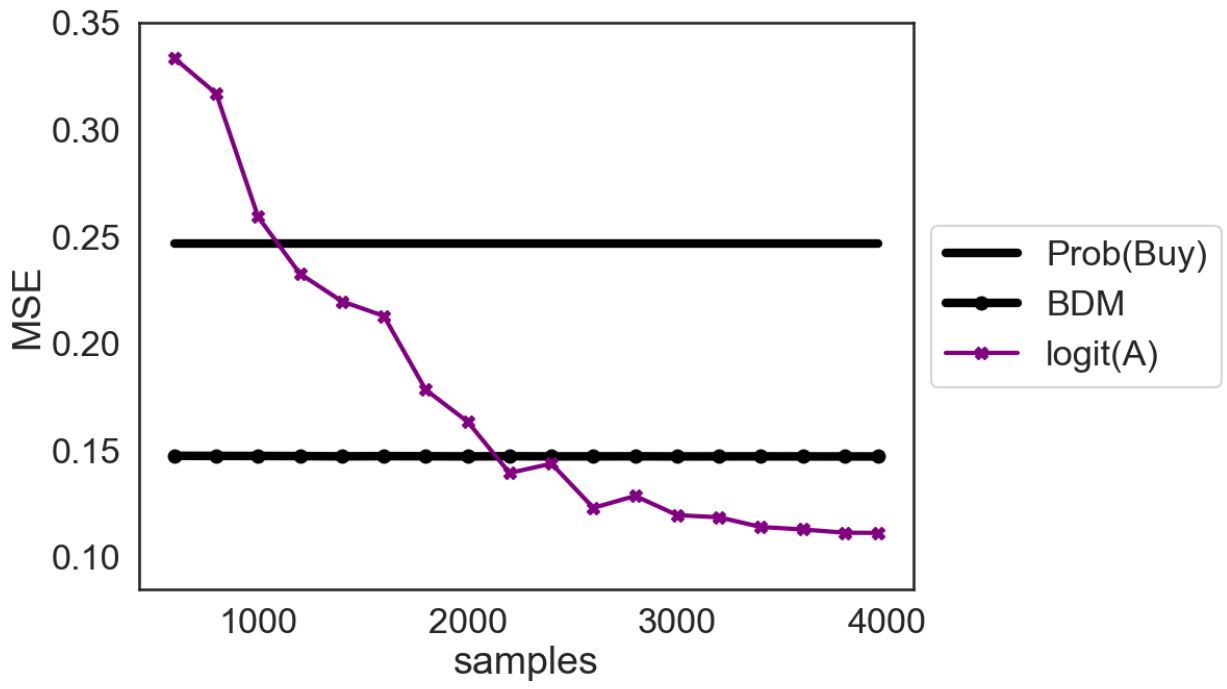

(b) Logit results with 2AFC data.

Notes: Out-of-sample MSE estimated on 440 hold-out observations. Sample size increases from 600 to 3960 in intervals of 200. Since samples are random, we repeat the estimation 50 times and MSE is averaged. This is a companion figure to Figure 4 in the main text.

Table A2: Test Sample — Holding Out Observations vs. Holding Out Subjects

| Models | Within&Between-Subjects MSE | Between-Subjects MSE | Between-Items MSE |
|---|---|---|---|
| BDM | 0.1473 (0.0019) | 0.1469 (0.0033) | 0.1445 (0.0025) |
| logit(C) | 0.1506 (0.0016) | 0.1866 (0.0019) | 0.1689 (0.0022) |
| logit(W) | 0.1121 (0.0015) | 0.1312 (0.0024) | 0.1190 (0.0018) |
| logit(WO) | 0.1315 (0.0023) | 0.1913 (0.0047) | 0.1190 (0.0018) |
| logit(A) | 0.1115 (0.0016) | 0.2928 (0.0121) | 0.5037 (0.0144) |
| logit(AR) | 0.1240 (0.0061) | 0.4341 (0.0126) | 0.4986 (0.0151) |
| logit(WA) | 0.1120 (0.0080) | 0.2434 (0.0105) | 0.4710 (0.0148) |
| logit(WOA) | 0.1784 (0.0023) | 0.4579 (0.0149) | 0.4876 (0.0124) |
| logit(WOAR) | 0.1915 (0.0068) | 0.4762 (0.0117) | 0.4947 (0.0158) |
| lasso(C) | 0.1570 (0.0014) | 0.1860 (0.0018) | 0.1722 (0.0021) |
| lasso(W) | 0.1162 (0.0011) | 0.1301 (0.0022) | 0.1205 (0.0018) |
| lasso(WO) | 0.1145 (0.0012) | 0.1408 (0.0035) | 0.1193 (0.0017) |
| lasso(A) | 0.1044 (0.0012) | 0.1762 (0.0044) | 0.1160 (0.0017) |
| lasso(AR) | 0.1057 (0.0011) | 0.1865 (0.0047) | 0.1185 (0.0018) |
| lasso(WA) | 0.0937 (0.0012) | 0.1331 (0.0032) | 0.0995 (0.0015) |
| lasso(WOA) | 0.0944 (0.0011) | 0.1466 (0.0043) | 0.0988 (0.0014) |
| lasso(WOAR) | 0.0943 (0.0012) | 0.1536 (0.0046) | 0.1040 (0.0016) |
| rf(C) | 0.1464 (0.0016) | 0.2289 (0.0031) | 0.1845 (0.0027) |
| rf(W) | 0.1077 (0.0014) | 0.1573 (0.0034) | 0.1224 (0.0027) |
| rf(WO) | 0.1013 (0.0012) | 0.1340 (0.0027) | 0.1248 (0.0023) |
| rf(A) | 0.1010 (0.0010) | 0.1435 (0.0026) | 0.1138 (0.0015) |
| rf(AR) | 0.1069 (0.0010) | 0.1443 (0.0025) | 0.1201 (0.0014) |
| rf(WA) | 0.0919 (0.0009) | 0.1226 (0.0021) | 0.0986 (0.0013) |
| rf(WOA) | 0.0890 (0.0009) | 0.1187 (0.0020) | 0.0970 (0.0013) |
| rf(WOAR) | 0.0927 (0.0010) | 0.1186 (0.0020) | 0.1008 (0.0013) |

Notes: Summary table of out-of-sample MSE results for all three statistical methods, using the full available sample of features. Standard errors across the 50 runs are reported in parentheses.

Figure A3: AUC: Performance of Supervised Machine Learning using WTP Data

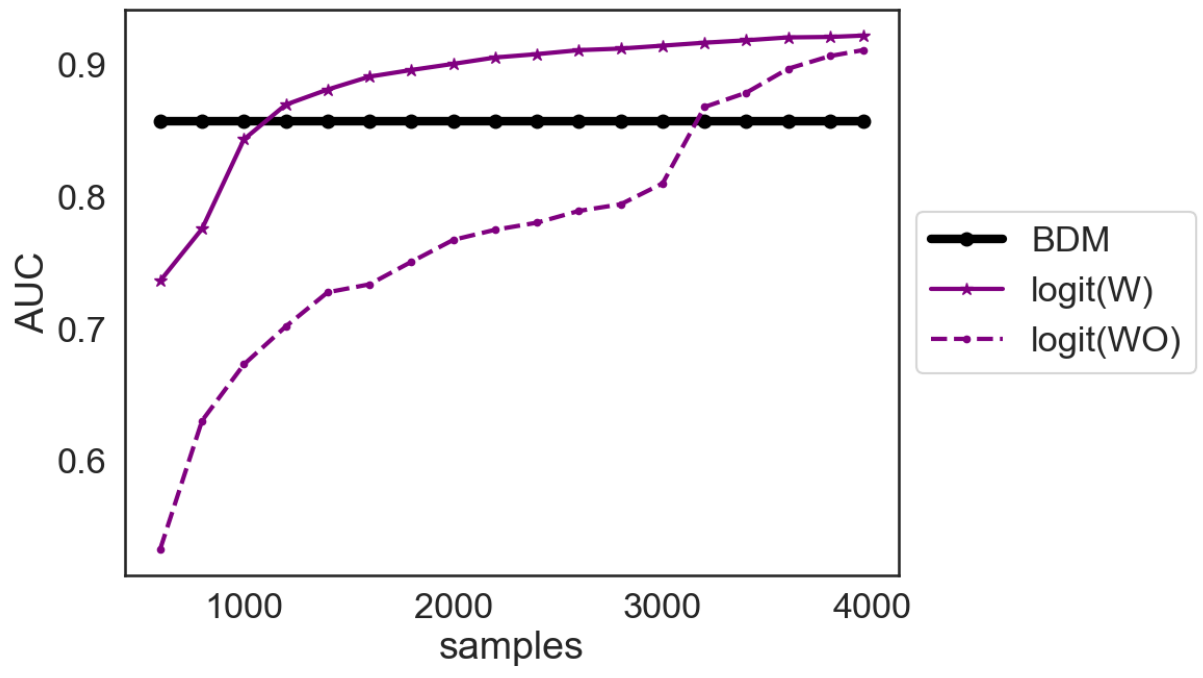

(a) Logit results with WTP data.

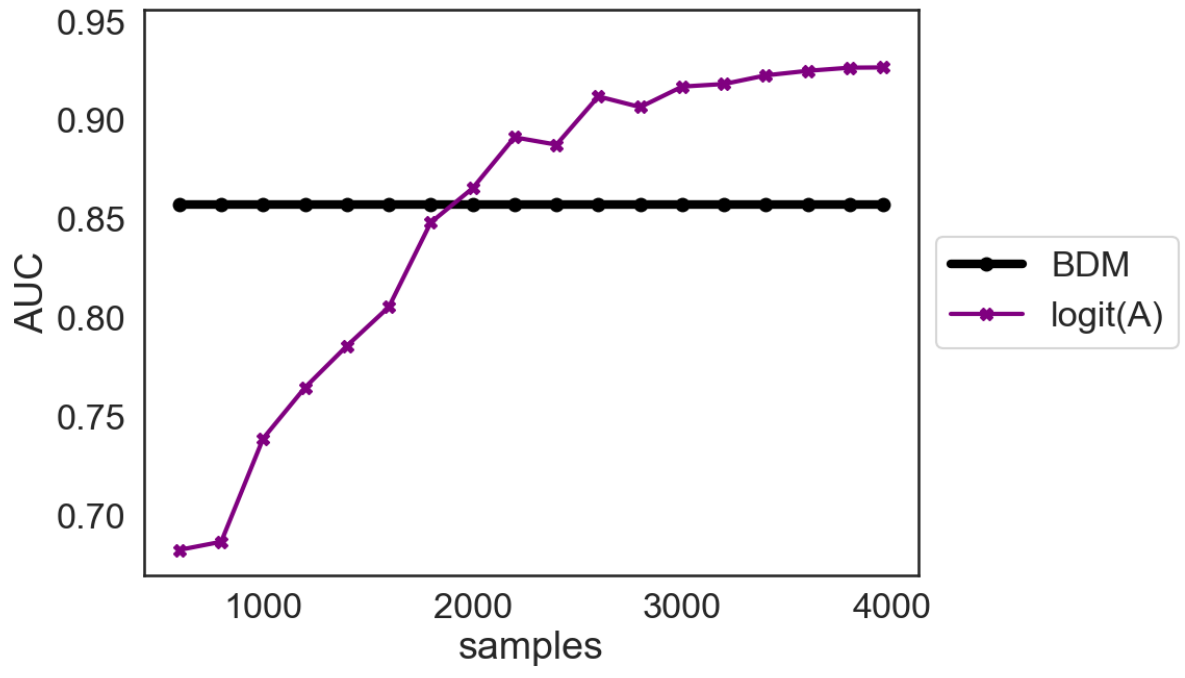

(b) Logit results with 2AFC data.

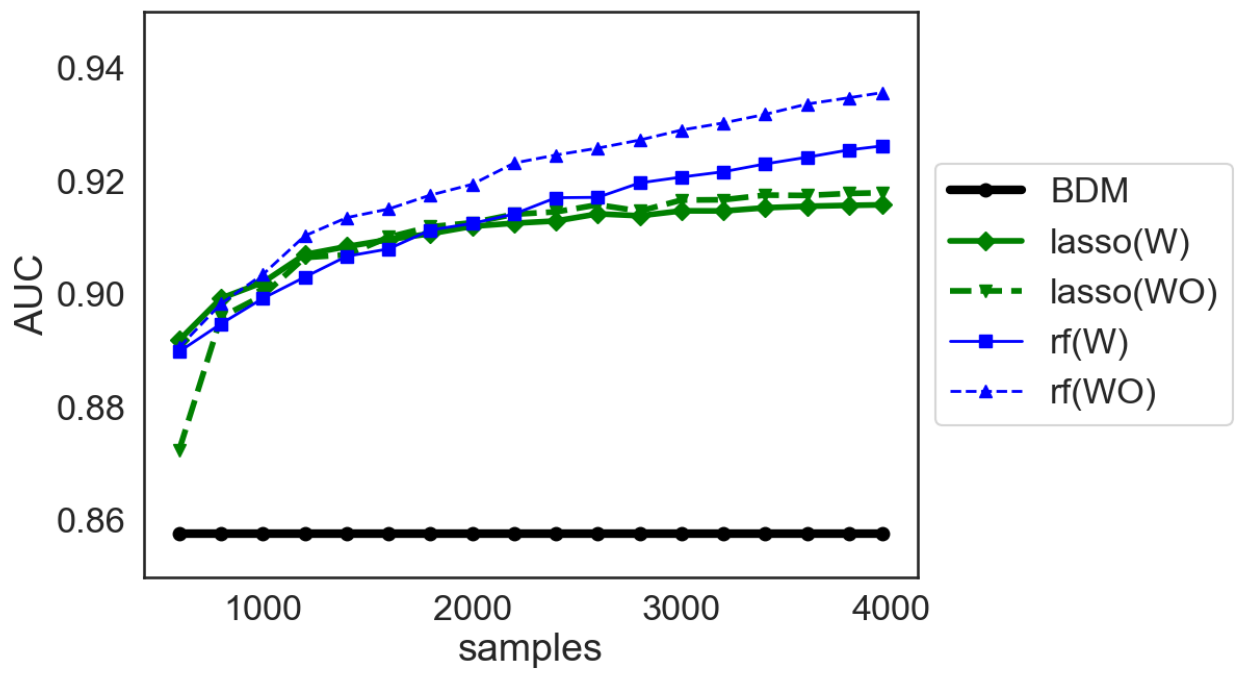

(c) Lasso and random forest results with WTP data.

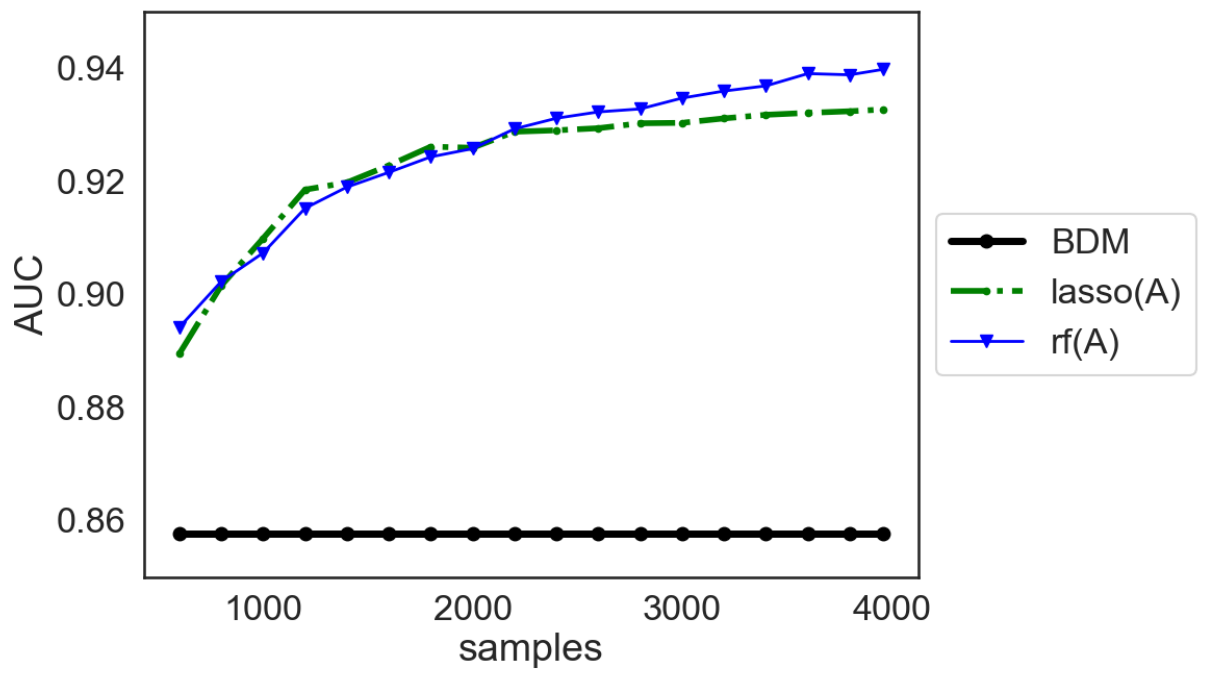

(d) Lasso and random forest results with 2AFC data.

Notes: Out-of-sample AUC estimated on 440 hold-out observations. Sample size increases from 600 to 3960 in intervals of 200. Since samples are random, we repeat the estimation 50 times and AUC is averaged.

Figure A4: Performance of Supervised Machine Learning with All Combinations of Data

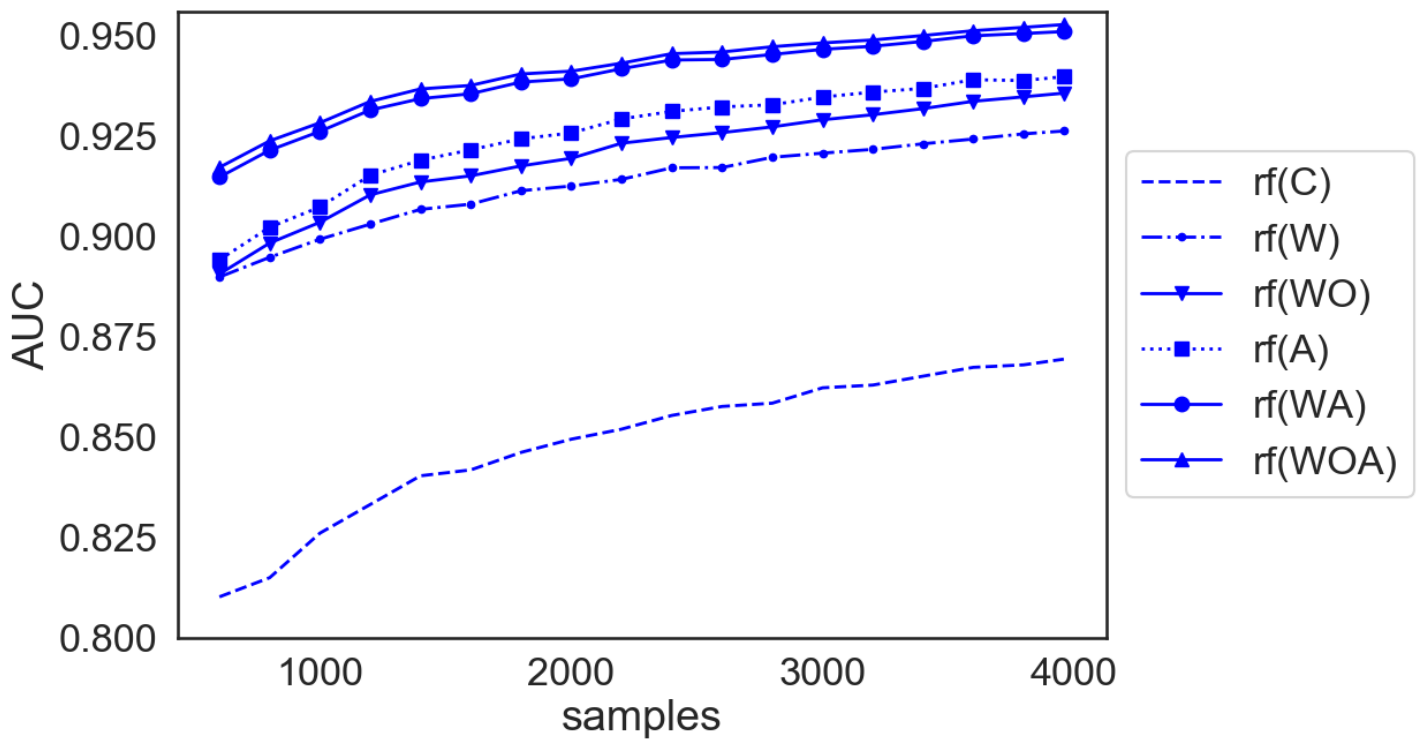

(a) Results for random forest algorithm.

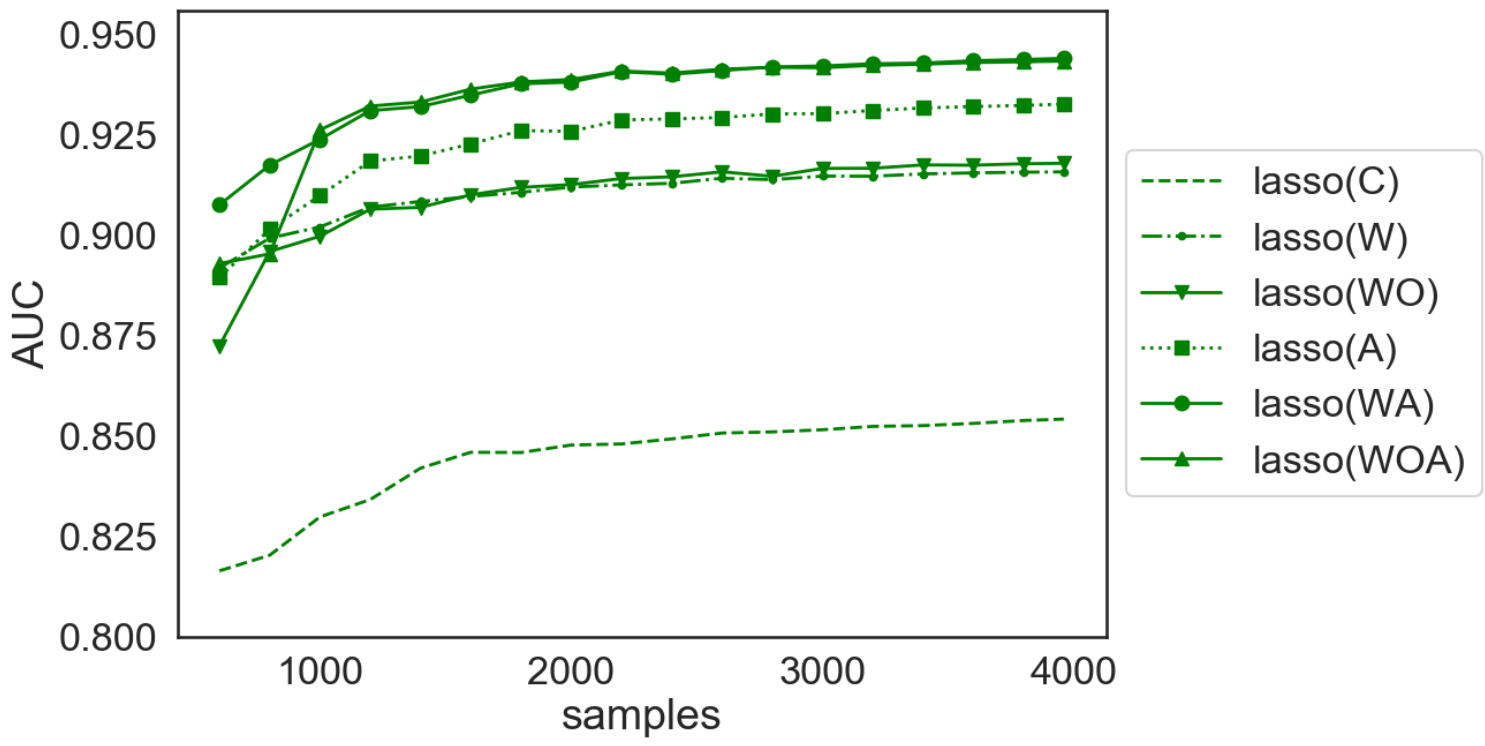

(b) Results for lasso algorithm.

Notes: Out-of-sample AUC estimated on 440 hold-out observations. Sample size increases from 600 to 3960 in intervals of 200. Since samples are random, we repeat the estimation 50 times and AUC is averaged.

Figure A5: Performance of Supervised Machine Learning with RT Data

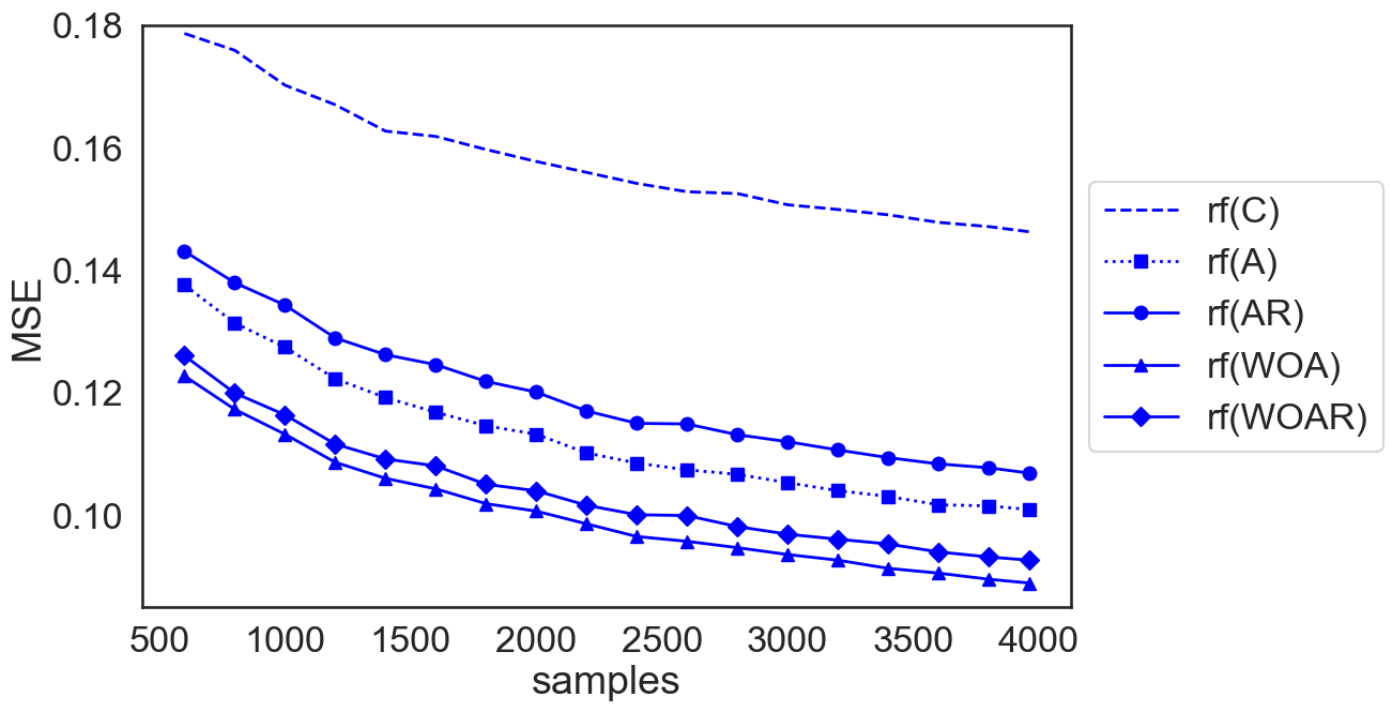

(a) Results for random forest algorithm.

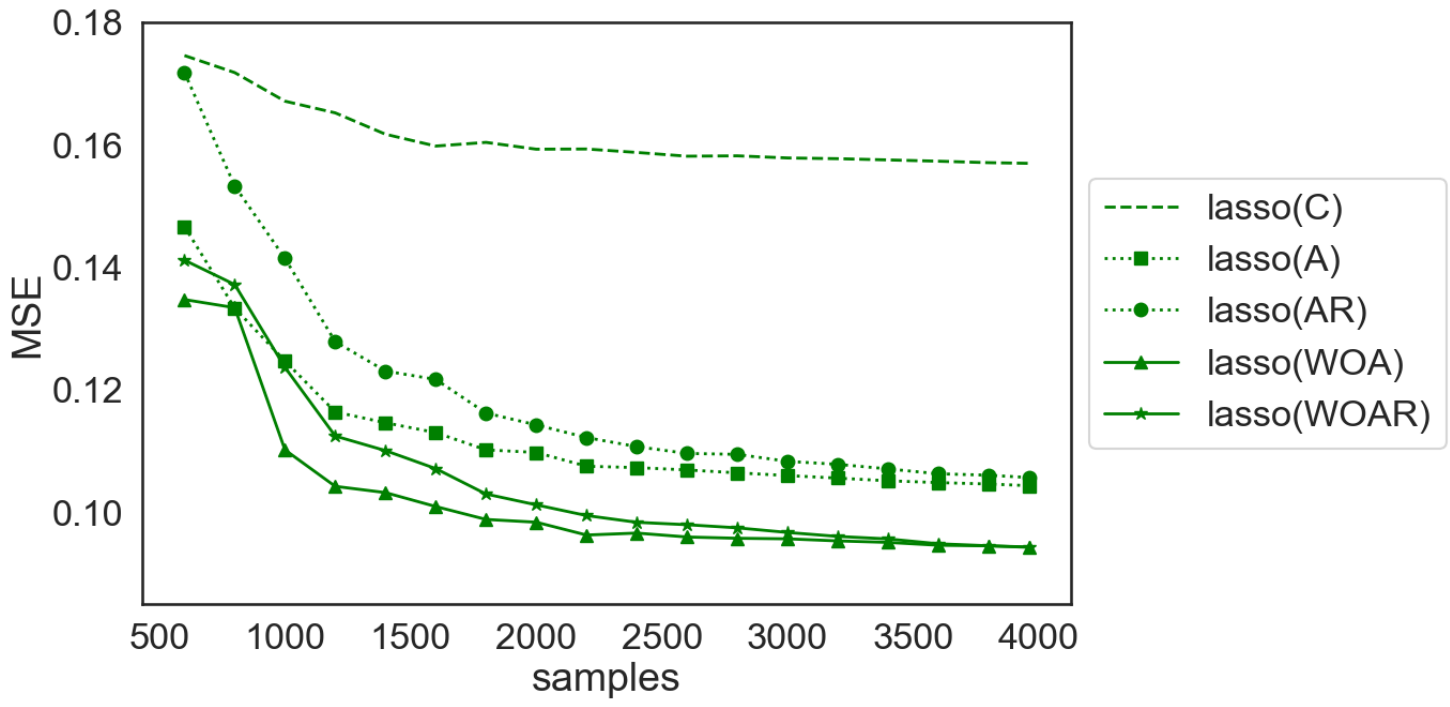

(b) Results for lasso algorithm.

Notes: Out-of-sample MSE estimated on 440 hold-out observations. Sample size increases from 600 to 3960 in intervals of 200. Since samples are random, we repeat the estimation 50 times and MSE is averaged.

# C Instrument

This section contains the text for instructions and the quiz given to subjects regarding the willingness-to-pay elicitation procedure.

Welcome. Thank you for participating in this experiment. This is an experiment about decision making. It should take approximately one hour.

At the end you will be paid your earnings in cash. By agreeing to participate, you have already earned a show-up fee of $7. You have also stated that you like snacks and do not have any food allergies. You will earn a total of $20, however this requires your participation through the entire experiment. Based on your choice, you may also receive a snack which you may pay for from your earnings. If you obtain a snack you will be given time at the end of the experiment to eat it. You have to have at least some of the snack before you leave.

Please make sure all of your personal belongings are below your desk. Please remain quiet for the rest of the experiment. If you have any questions, wait until the end of the instructions and ask.

The next few pages provide detailed instructions about the experiment. There is no deception in the experiment - we will do everything as outlined in the instructions.

There are four tasks in this experiment. The experimenter will read the instructions for each task just before you begin the task. One trial from the second, third, or fourth task will be randomly chosen to count for real stakes. On this trial your choice will determine what payoff and food you obtain.

You are welcome to ask any clarifying questions about the tasks or about the procedures.

**Part I**

In this task, you will designate items at random to be your bonus items. There are two bonus items: Gold and Silver ones. Throughout the experiment, if you ever obtain your Gold item you will receive an additional $4, and if you ever obtain your Silver item you will receive an additional $2.

An example trial is shown below. You will see 20 cards on the screen, and each card has one of 20 snacks. First, you choose one of 20 cards by ‘left clicking your mouse button’ on the card. Second, you choose another card from 19 remaining cards. The items revealed will now be your bonus items. In this example, Milky Way is your Gold item, and Whoppers is your Silver item. During the experiment, you will see a gold border around the Gold item and a silver border around the Silver item in all subsequent tasks. When you evaluate the Gold item, indicated by the gold border, don’t forget you get 4 additional dollars whenever you obtain the item. Similarly, you should not forget you get 2 additional dollars when you evaluate the Silver item with the silver border.

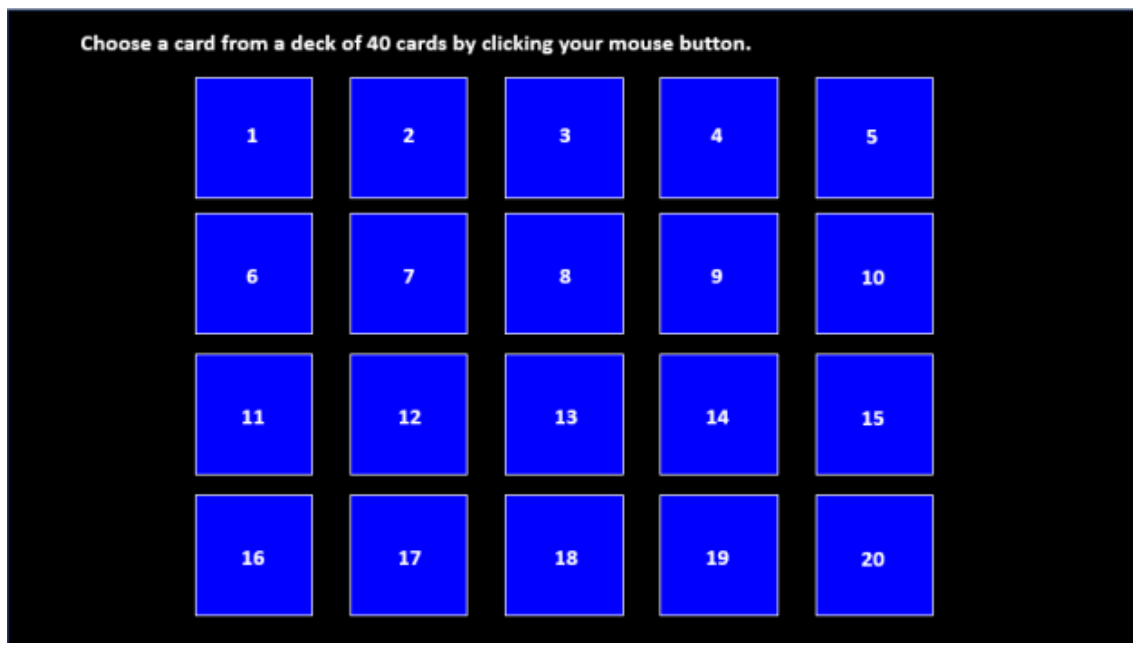

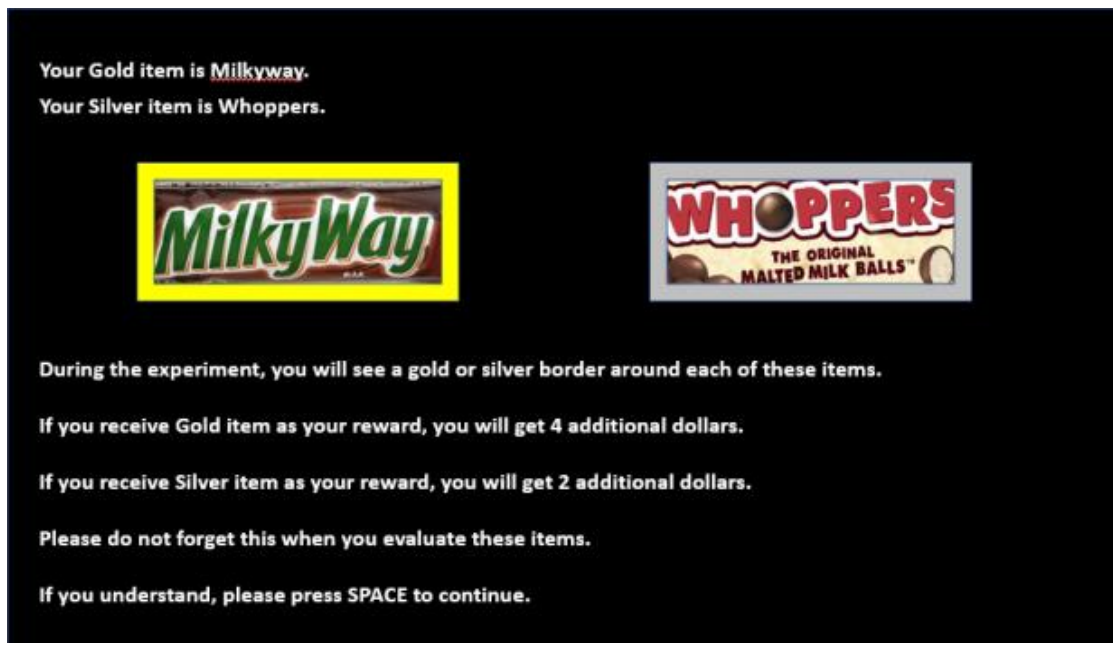

The basic steps to keep in mind in this task are:

1. You will see 20 cards on the screen.
2. You will choose one of 20 cards by left clicking on the card.
3. You will choose one more card from 19 cards by left clicking on the card.
4. This will reveal your bonus items which will be henceforth indicated by a gold or silver border.
5. If you ever obtain the Gold item, you will receive an additional $4.
6. If you ever obtain the Silver item, you will receive an additional $2.

**Do you have any questions? Please raise your hand and ask any questions.**

**Part II**

In this task, you will be asked about your willingness-to-pay for each snack. For example, a single trial will ask how much you are willing to pay to eat a Milky Way. Each snack is approximately a single serving.

In each trial, you will be shown an image of a snack. If Milky Way is your Gold item it will have a gold border. You can answer your willingness-to-pay by left clicking your mouse button on the monetary values at the bottom of the screen. You can choose your willingness-to-pay for that item from $0 to $5.75 in increments of $0.25.

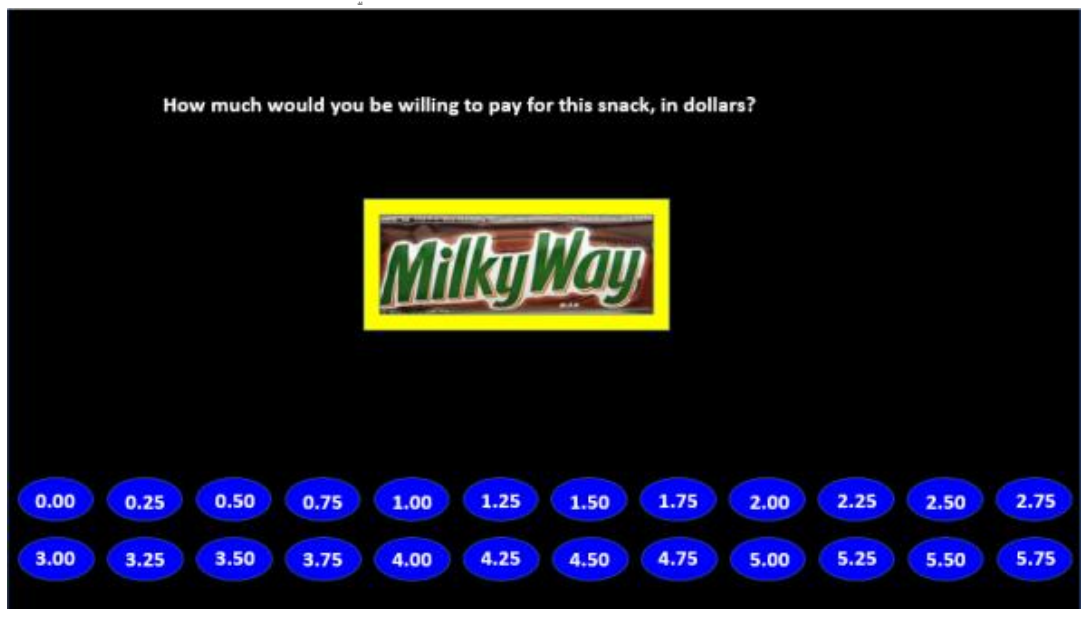

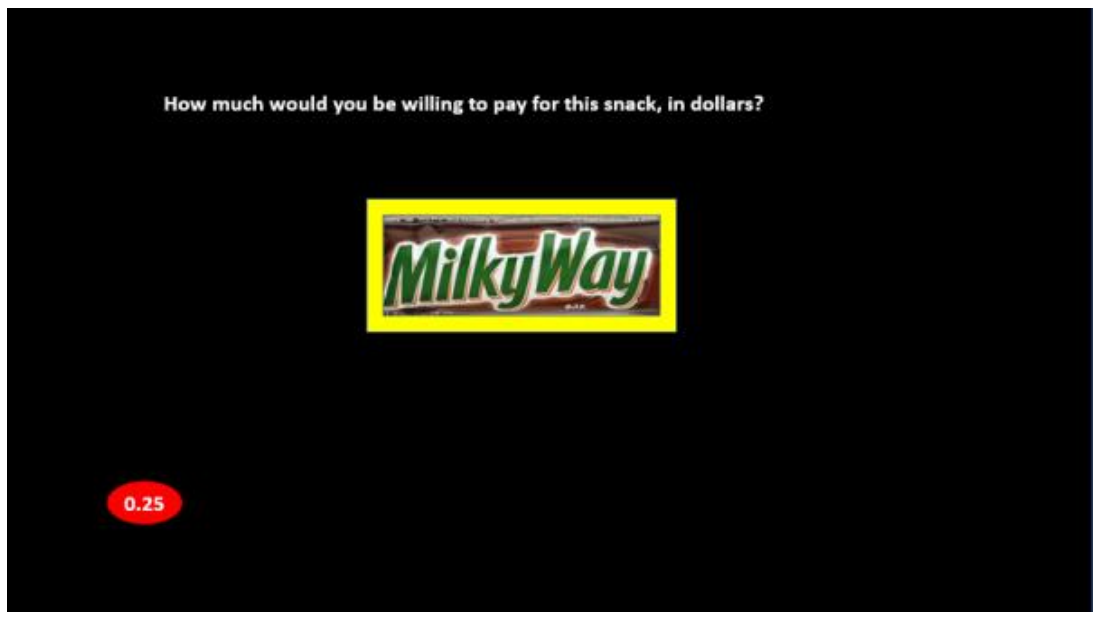

The screen will show you the amount you select. In this example above, $0.25 is selected as the willingness-to-pay for Milky Way.

The computer will randomly select a price. You will obtain the snack as long as the random price is equal to or below your stated willingness-to-pay. Notice that it is optimal to state the maximum of what you would pay for this item. It is as if a friend was going to the store and asked you if you would like him to pick up a snack for you. There is a snack you want but you don't know the price. What would you do? You would inform your friend to only buy the snack so long as the price is equal to or below your true valuation of the snack. This task is logically equivalent. The procedure ensures that it is best for you truthfully reveal the highest price you are willing to pay.

As an example, imagine your true willingness-to-pay for a Dr. Pepper is $3.00, but you untruthfully overstate that your willingness-to-pay is $4.00. Then, if the random price is $3.50, you would pay $3.50 and get the Dr. Pepper even though the most you would be willing to pay is $3.00. You would have wasted $0.50. If the price were below $3 or above $3.50 the outcome would have been the same as if you reported your true valuation. So you can only do worse if you give a willingness-to-pay above your true valuation. In contrast, if you understate your willingness-to-pay as $2.00, and if the random price is $2.50, you would be disappointed because you would not buy the Dr. Pepper even though the price is below your true valuation. If the price were below $2.00 or above $3.00 the outcome would have been the same as if your reported your true valuation. So you can only do worse if you give a willingness-to-pay below your true valuation.

Note that you cannot influence the purchase price with your stated willingness-to-pay, because the purchase price is completely random and independent of whatever you state.

The situation is analogous to having a friend go to the store to buy items on your behalf, but in which you don't know the prices. The optimal thing to do is to tell your friend to buy the Dr. Pepper only if the price is at your willingness to pay or lower ($3.00). Saying anything else would have your friend not buying when you would have wanted it or buying at a price too high.

However, if you state willingness-to-pay for your Gold item, please remember your bonus item comes with $4 in cash. Thus your Gold item should be worth whatever the value of that item is to you plus $4.

Likewise, if you state willingness-to-pay for your Silver item, please remember your bonus item comes with $2 in cash. Thus your Silver item should be worth whatever the value of that item is to you plus $2.

Finally, you should treat every trial as if it is the only one that matters since only one trial will be chosen at random to count for real stakes. Please take each and every decision seriously.

The basic steps to keep in mind in this task are:
1. You will see a snack on the screen.

2. You should answer your maximum willingness-to-pay by left clicking your mouse button on one of the amounts from $0 to $5.75 in increments of $0.25.
3. You are best off by selecting the maximum you would be willing to pay.
4. When you evaluate each snack, please do not forget you will get an additional $4 for your Gold item and an additional $2 for your Silver item.
5. You should treat every trial as if it is the only one that matters because you don’t know which trial will be chosen.

**Do you have any questions? Please raise your hand or ask the experimenter any questions you have so far. The experimenter will tell you when you can start the experiment.**

**Quiz for Part II.**

DATE___________ SUBJECT__________

Please answer the following four questions by circling the correct answer:

1.Imagine your true value for a Coke is $2.00. What happens if you answer your willingness-to-pay for it is $3.00?

1) You have no risk of paying more for the Coke than your true value.
2) You can influence the random price of a Coke by overstating your true value for it.
3) It's possible that you would pay a higher price than your true value for a Coke to receive it.

2.Imagine your true value for a Sprite is $2.75. What happens if you answer your willingness-to-pay for it is $2.00?

1) Your chance of receiving the Sprite is the same.
2) You lose the chance to get a Sprite if the random price is between $2.00 and $2.75.
3) You can strategically influence the random price of a Sprite by understating your true value for it.

3. Which of the following two statements is true?

1) You state your willingness-to-pay is $2.00 and the random price is $2.50. You receive a snack and pay $2.50 for it.
2) You state your willingness-to-pay is $2.00 and the random price is $1.50. You receive a snack and pay $1.50 for it.

4. Which of the following two statements is true?

1) You state your willingness-to-pay is $3.25 and the random price is $2.00. You do not receive the snack and you pay nothing.
2) You state your willingness-to-pay is $3.25 and a random price is $4.00. You do not receive the snack and you pay nothing.

5. Suppose MilkyWay is worth $1.50 to you and it is your Gold item. So, if you were to receive a MilkyWay you would also get the cash bonus. Which of the following bids makes you the best off?"

(a) $1.50 (b) $3.50 (c) $5.50

6. Suppose Dr.Pepper is worth $2.25 to you and it is your Silver item. Which of the following bids makes you the best off?"

(a) $2.25 (b) $4.25 (c) $6.25

**Please raise your hand to turn in your quiz when you are finished.**

**Part III**

In this task, you will make decisions about which of two possible snacks to consume. For example, a single trial will ask whether you would you prefer to eat Milky Way or Whoppers. Each of the snacks for this task is approximately a single serving. Over the course of the task, you will see each snack item several times.

In each trial, you will be shown two images, as in the example below. You will respond with your left and right index fingers, one for each button. You will only need to press the ‘c’ or ‘m’ button. Importantly, please use both of your hands in this task. The options on the left can be chosen with the ‘c’ button and the option on the right can be chosen with ‘m’ button.

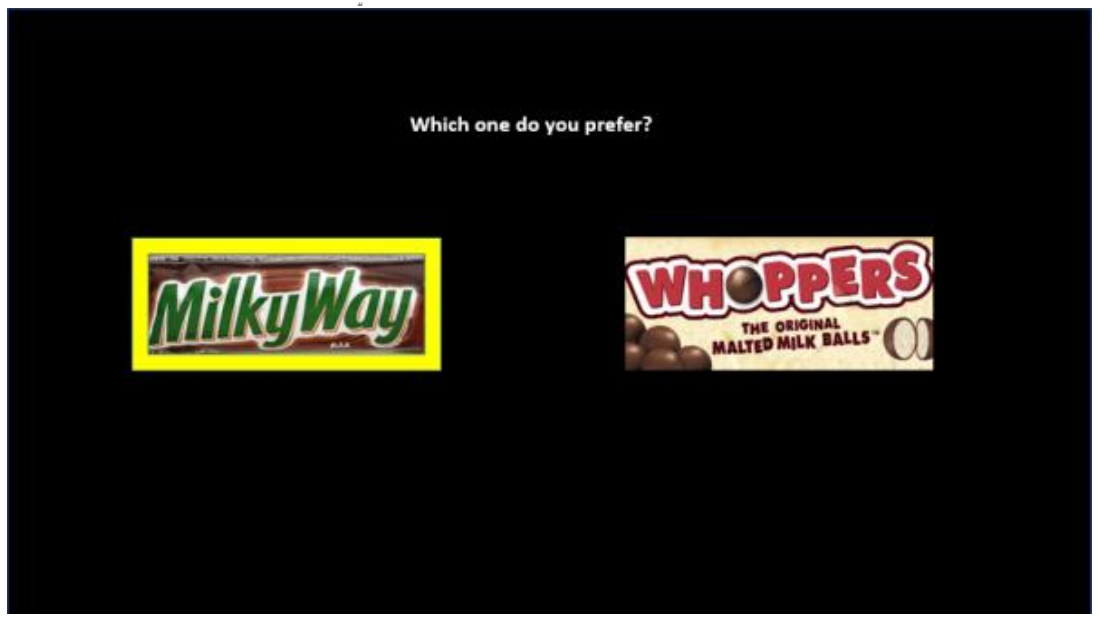

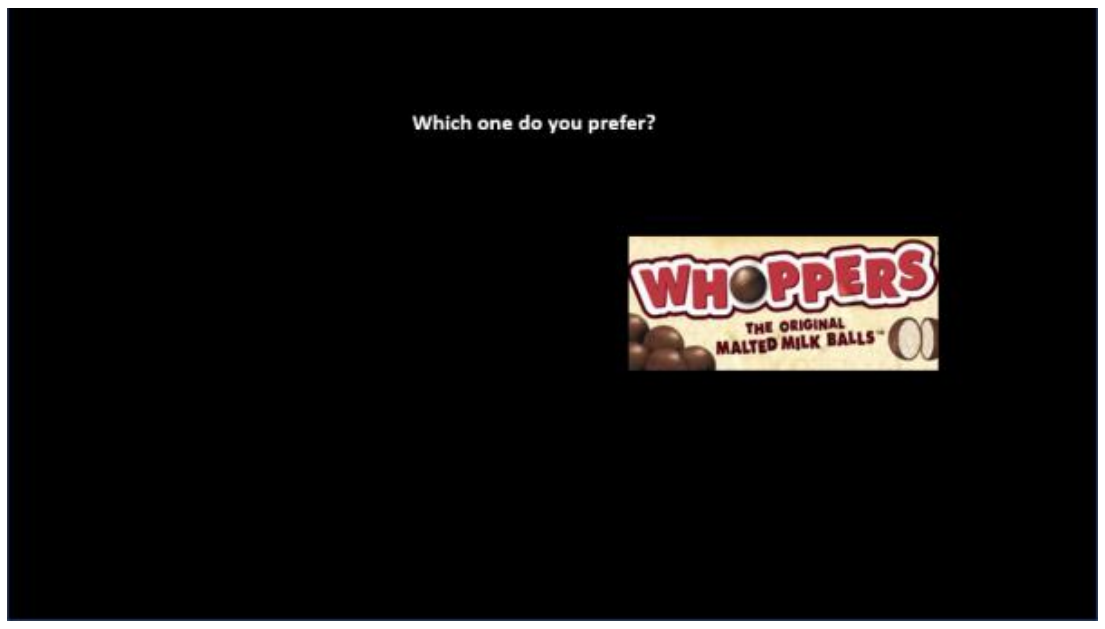

The screen will show you the option you select. The example above shows the screen in which the choice is between Milky Way and Whoppers, with Whoppers being selected.

Since you don’t know which trial will be chosen, you should treat every trial as if it is the only one that matters. Please treat each decision as a real choice.

The basic steps to keep in mind in this task are:

1. You will be shown a picture of two snack options.
2. Please use both index fingers to select which snack you prefer.
3. Again, please don’t forget you will get the cash bonus for Gold or Silver items.

**Do you have any questions? Please raise your hand or ask the experimenter any questions you have so far. The experimenter will tell you when you can start the experiment.**

**Part IV**

In this task, you will be asked whether you will buy each snack at a given price. For example, the screenshot below presents the choice to buy Milky Way for $1.00. Over the course of the task, you will see each snack several times. Each of the snacks is approximately a single serving.

You will respond with your left and right index fingers, one for each button. You will only need to press the ‘c’ or ‘m’ button. Importantly, please use both of your hands in this task. To some participants, the options YES and NO will be on the opposite side. (No on the left and YES on the right.) The options on the left can be chosen with the ‘c’ button and the option on the right can be chosen with ‘m’ button.

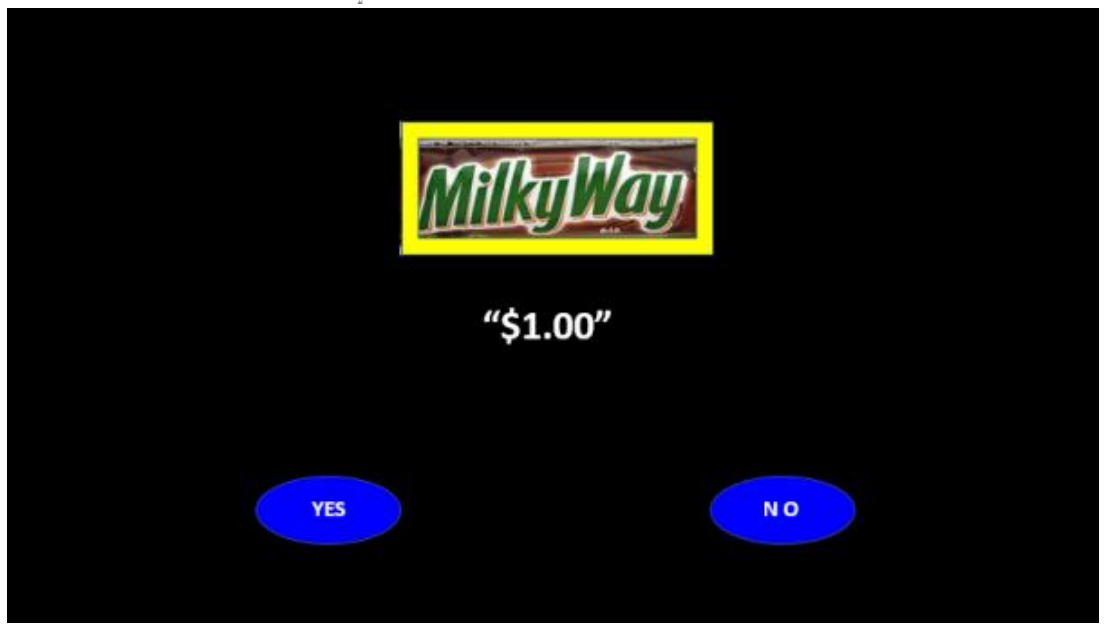

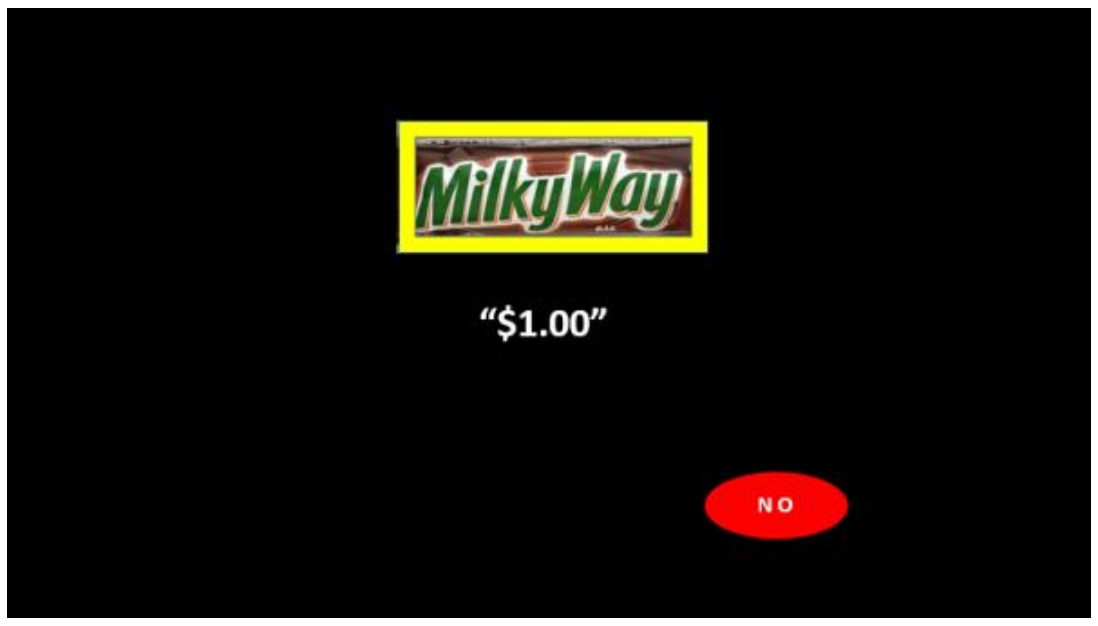

Since you don’t know which trial will be chosen, you should treat every trial as if it is the only one that matters.

The basic steps to keep in mind in this task are:
1. You will be shown a picture of a snack on the screen.
2. You will answer whether you would buy that snack at the given price.
3. By using both index fingers, please press ‘c’ if your answer is on the left, and press ‘m’ if your answer is on the right.
4. It will be best for you to respond honestly, and treat each decision as a real choice.
5. Again, please don’t forget you will get the cash bonus for Gold or Silver items.

**Do you have any questions? Please raise your hand or ask the experimenter any questions you have so far. The experimenter will tell you when you can start the experiment.**

## Screenshots

This section contains several screen shots from the tasks and payment procedure.

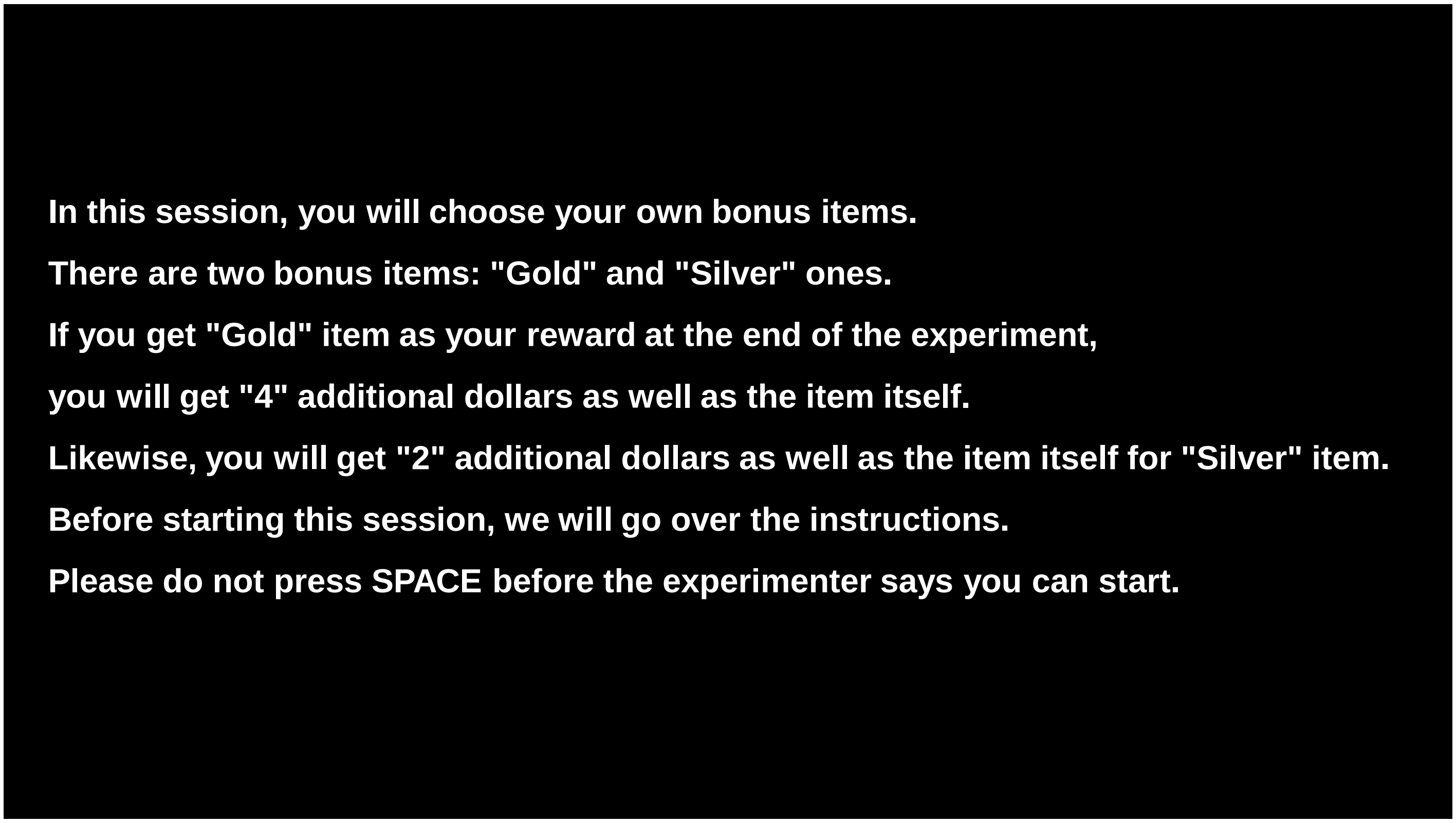
In this session, you will choose your own bonus items.
There are two bonus items: "Gold" and "Silver" ones.
If you get "Gold" item as your reward at the end of the experiment,
you will get "4" additional dollars as well as the item itself.
Likewise, you will get "2" additional dollars as well as the item itself for "Silver" item.
Before starting this session, we will go over the instructions.
Please do not press SPACE before the experimenter says you can start.

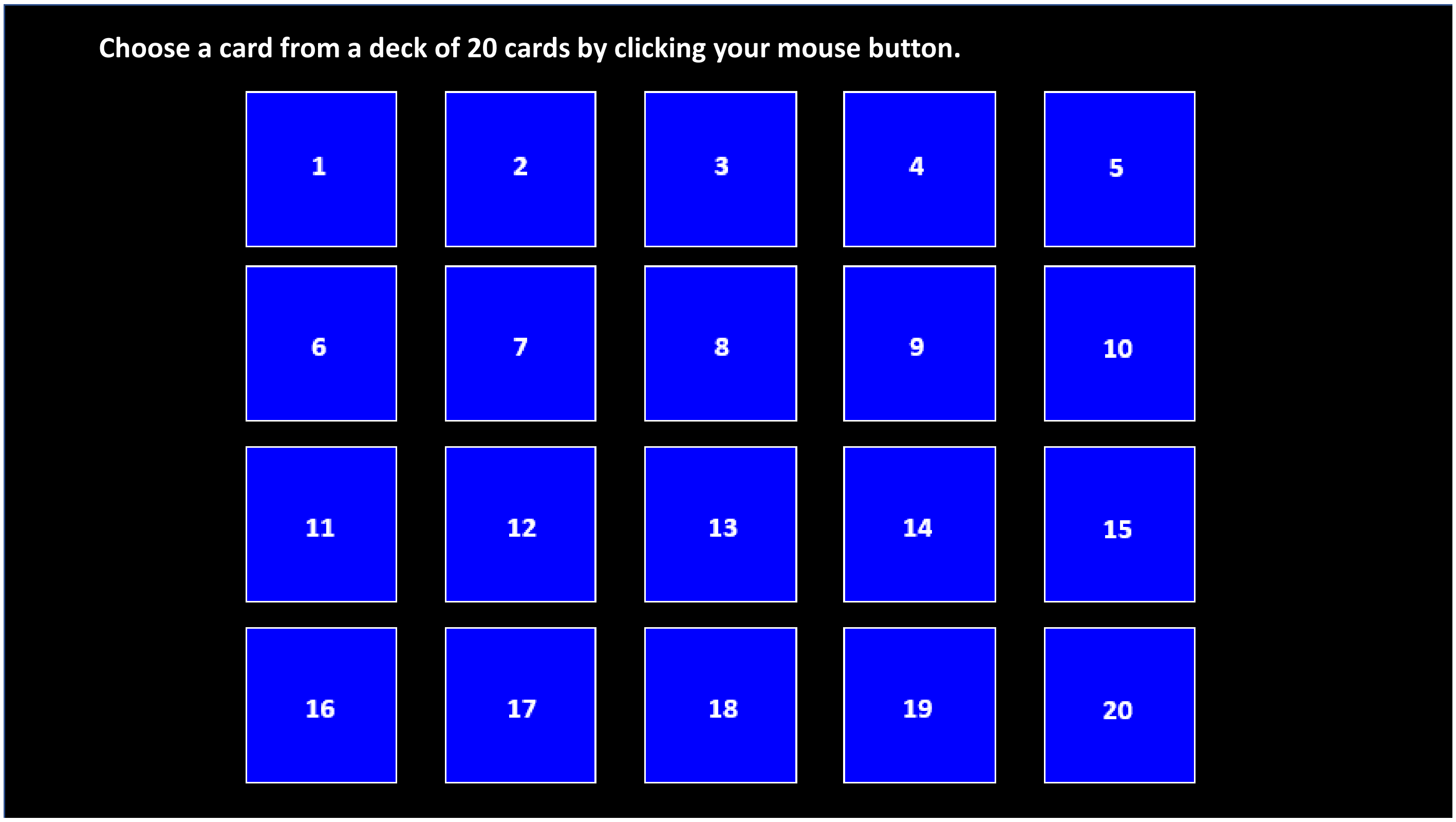
Choose a card from a deck of 20 cards by clicking your mouse button.
1
2
3
4
5
6
7
8
9
10
11
12
13
14
15
16
17
18
19
20

Choose one more card from a deck of 19 cards by clicking your mouse button.

| | | | | |
|---|---|---|---|---|
| 1 | 2 | 3 | 4 | 5 |
| 6 | 7 | 8 | 9 | 10 |
| 11 | 12 | 13 | 14 | 15 |
| 16 | 17 | 18 | 19 | 20 |

Your Gold item is Hershey's.

Your Silver item is Justins Peanut Butter.

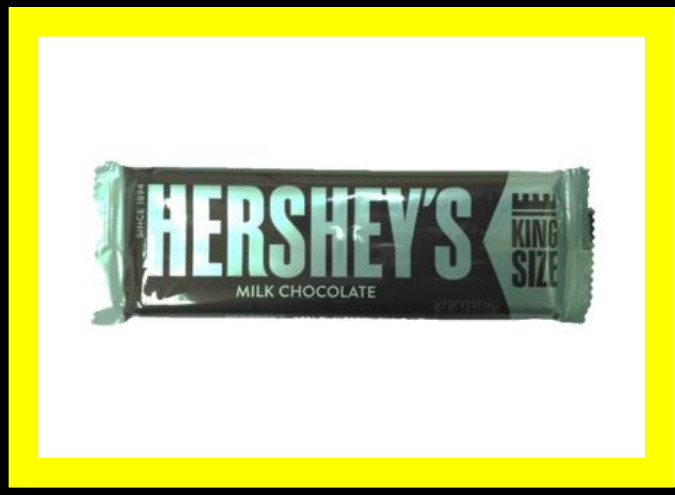

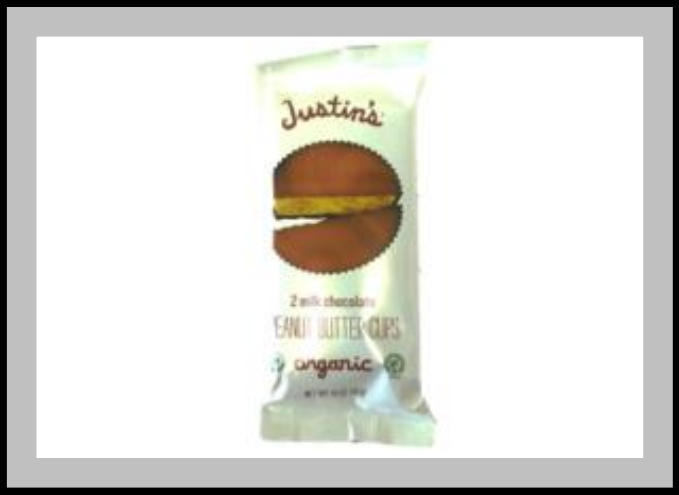

During the experiment, you will see a gold or silver border around each of these items.

If you receive Gold item as your reward, you will get 4 additional dollars.

If you receive Silver item as your reward, you will get 2 additional dollars.

Please do not forget this when you evaluate these items.

If you understand, please press SPACE to continue.

This is the end of the bonus item selection.

Thank you very much!

Please do not press SPACE before the experimenter says you can start.

In this session, you will see a picture of a snack several times.

You should answer how much you are willing to pay for it.

Please left click your mouse on the corresponding value.

Before starting this session, we will go over the instruction.

Please do not press SPACE before the experimenter says you can start.

How much would you be willing to pay for this snack, in dollars?

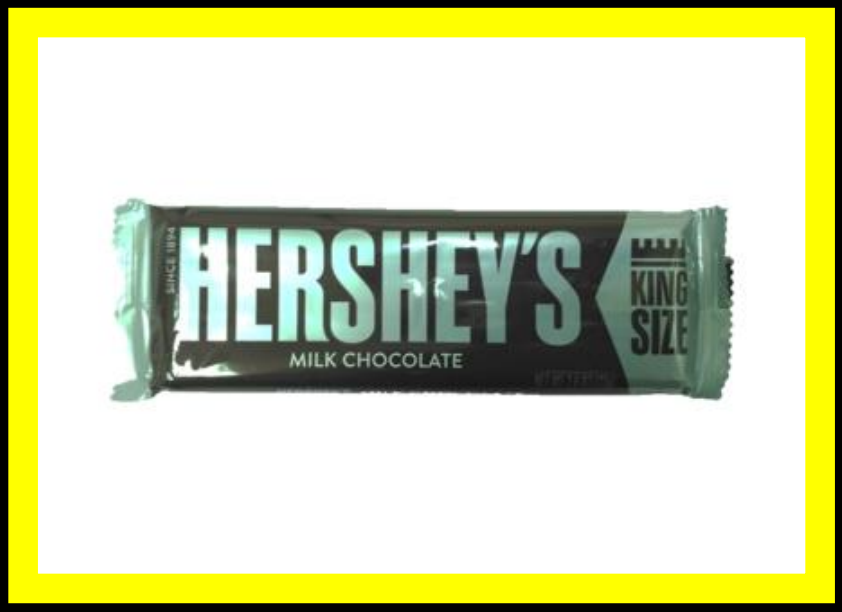

| | | | | | | | | | | | |
|---|---|---|---|---|---|---|---|---|---|---|---|
| 0.00 | 0.25 | 0.50 | 0.75 | 1.00 | 1.25 | 1.50 | 1.75 | 2.00 | 2.25 | 2.50 | 2.75 |
| 3.00 | 3.25 | 3.50 | 3.75 | 4.00 | 4.25 | 4.50 | 4.75 | 5.00 | 5.25 | 5.50 | 5.75 |

How much would you be willing to pay for this snack, in dollars?

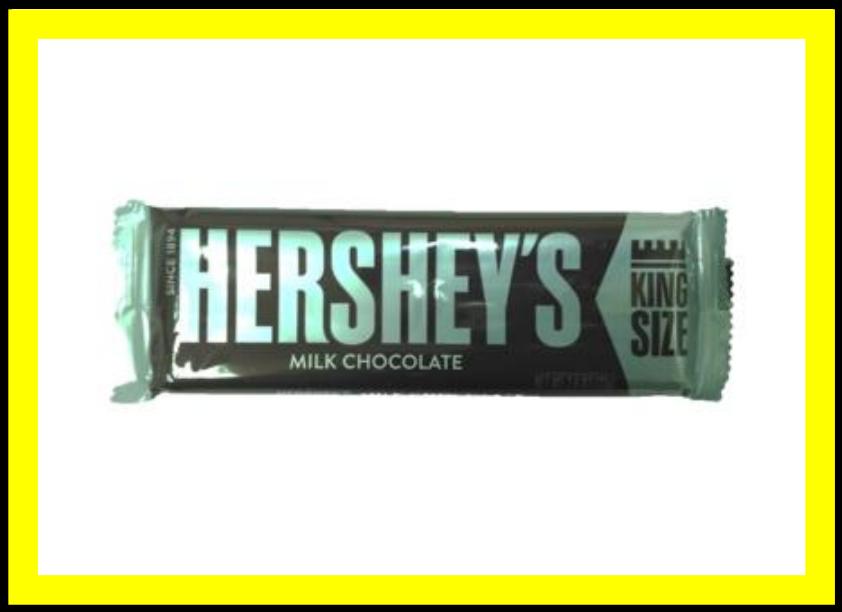

1.25

This is the end of the task.

Please raise your hand when you finish.

Thank you very much!

Each trial will show two snacks.

Please pick the one you prefer.

If you prefer the left item, press "c" on the keyboard.

If you prefer the right item, press "m" on the keyboard.

Remember to please use both index fingers to respond.

Before starting this session, we will go over the instruction.

Please do not press SPACE before the experimenter says you can start.

Which one do you prefer?

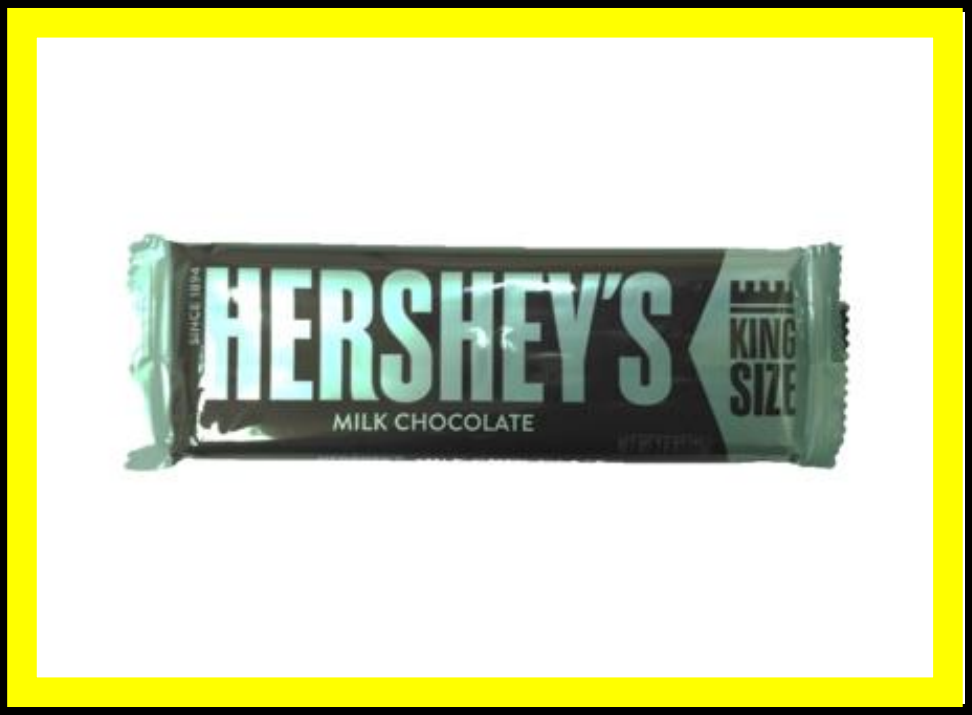

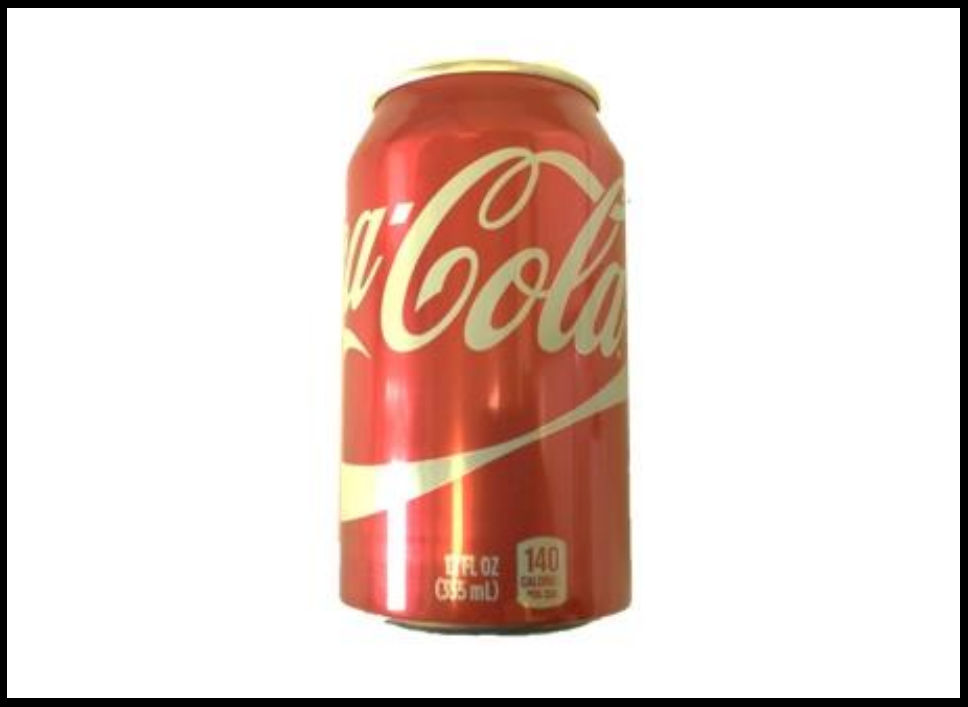

Which one do you prefer?

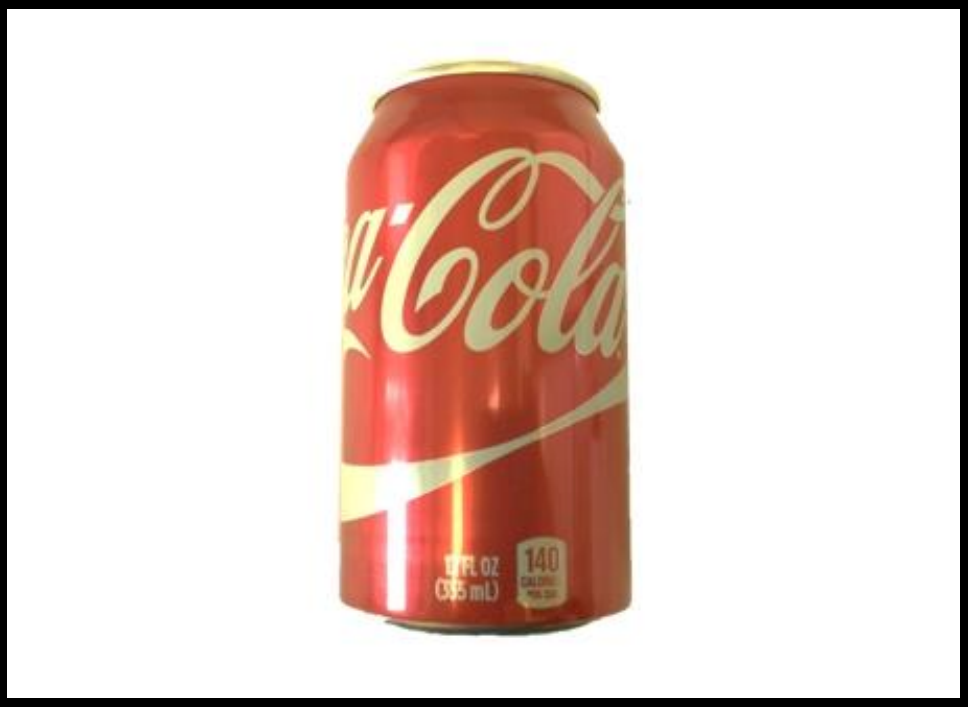

This is the end of the task.

Please raise your hand when you finish.

Thank you very much!

You will see one of 20 different snacks several times.

In each trial, you should decide whether to buy a snack for a given price.

If your answer is on the left, press "c" on the keyboard.

If your answer is on the right, press "m" on the keyboard.

Before starting this session, we will go over the instructions.

Please do not press SPACE before the experimenter says you can start.

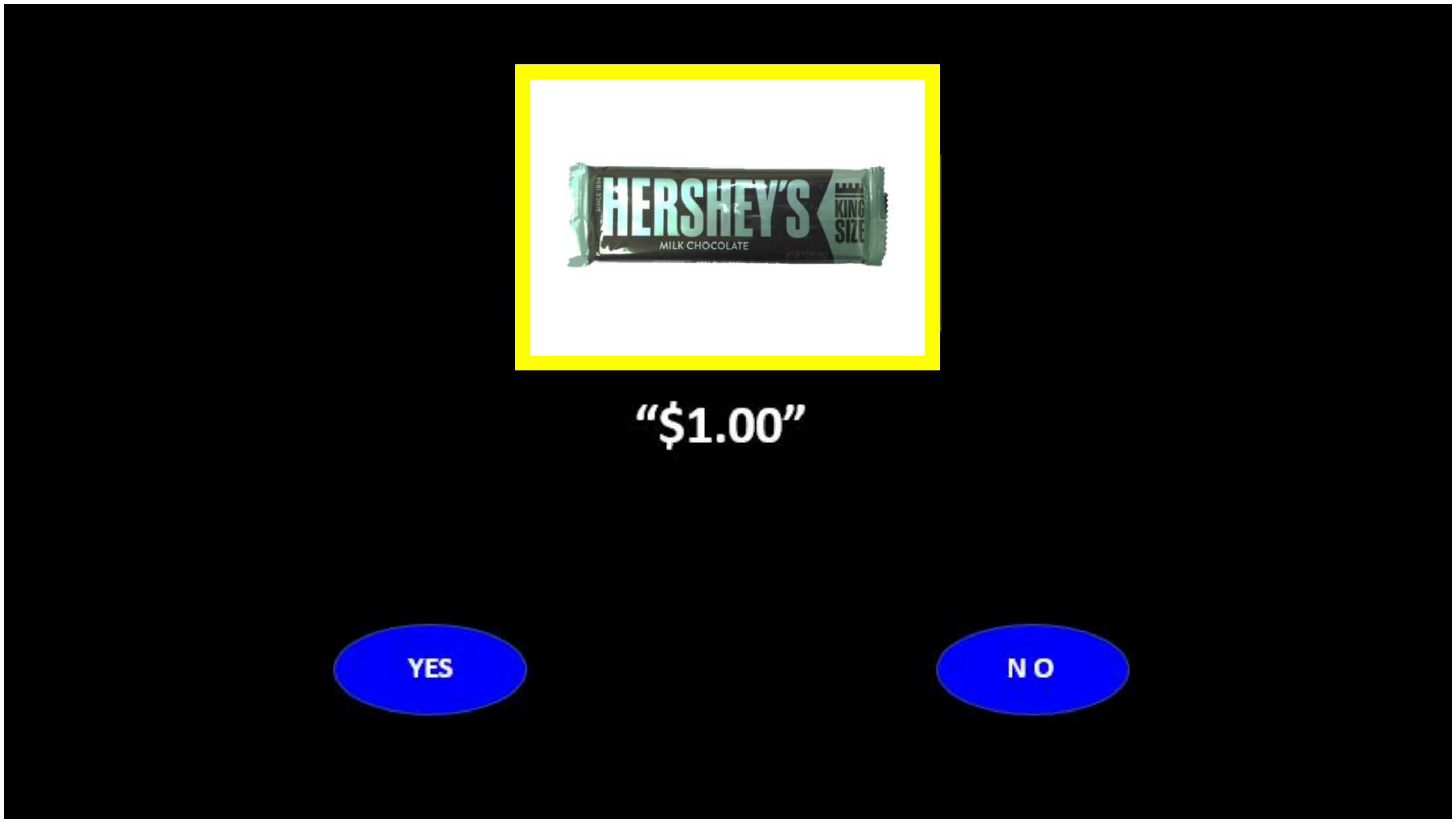

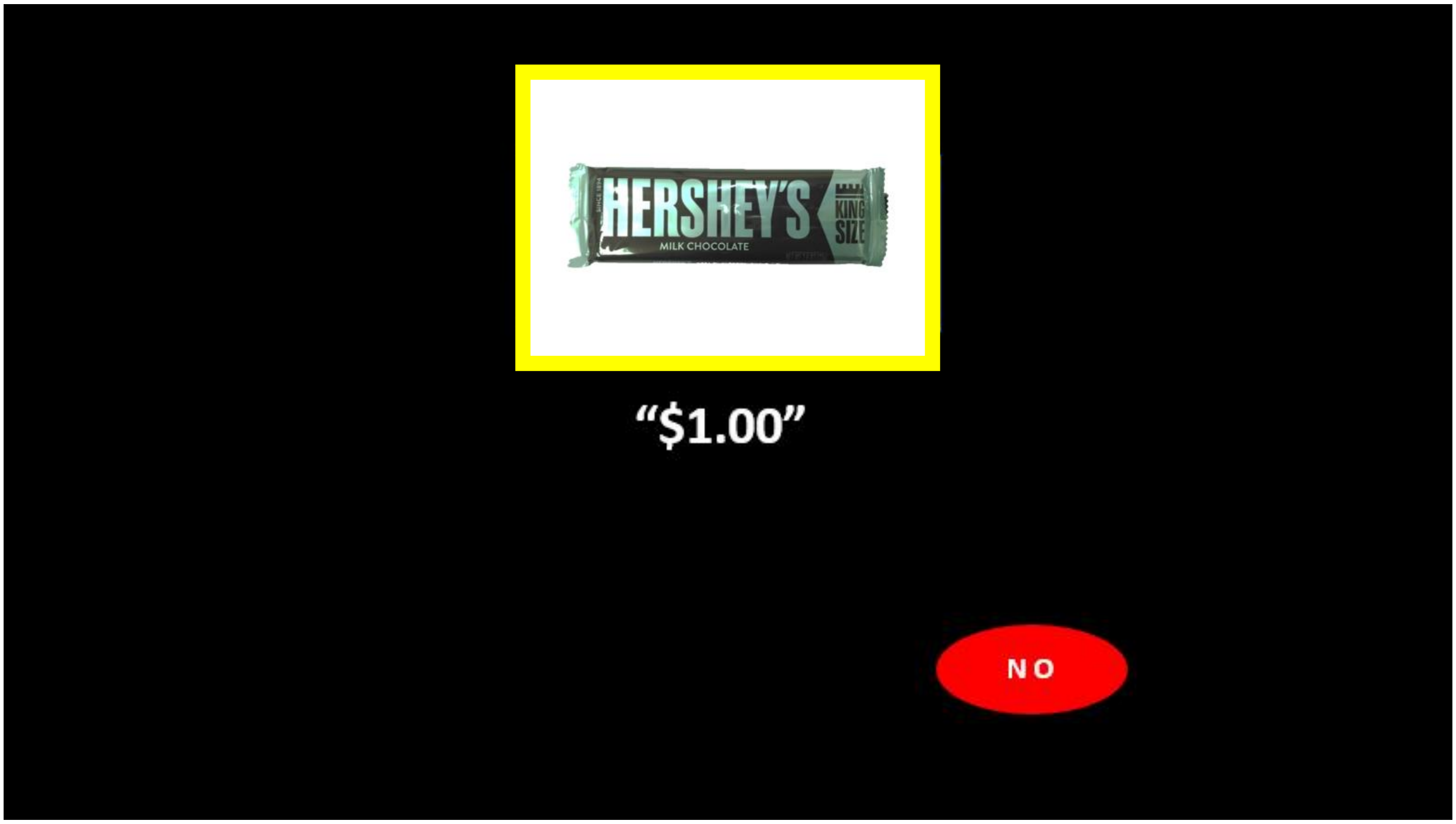

**This is the end of the task.**

**Please raise your hand when you finish.**

**Thank you very much!**

Now you will get a reward for this experiment.
Please roll the dice in front of you.

If you get 1 or 2, you will receive the reward on your willingness-to-pay for a snack in the 1st session.
If you get 3 or 4, you will receive the reward on your binary choice in the 2nd session.
If you get 5 or 6, you will receive the reward on your buy/no buy decision in the 3rd session.

Which number did you get on your dice?: 3

You will get a reward on a binary choice in the 2nd session.

The decision selected was between the Bar Cliff Peanut Crunch and the Pringles Original.
You chose the Pringles Original.
You will receive your food item now.
You must eat it before you leave the experiment.

THANK YOU FOR YOUR PARTICIPATION!